\documentclass[lettersize,journal]{IEEEtran}
\usepackage{amsmath,amsfonts}
\usepackage{algorithmic}
\usepackage{algorithm}
\usepackage{array}
\usepackage{booktabs}
\usepackage[caption=false]{subfig}
\usepackage{textcomp}
\usepackage{stfloats}
\usepackage{url}
\usepackage{color}
\usepackage{verbatim}
\usepackage{graphicx}
\usepackage{cite}
\usepackage{epstopdf}
\usepackage{graphics}
\usepackage{ntheorem}
\usepackage{theorem}
\usepackage{subfloat}
\allowdisplaybreaks[4]
\begin{document}

\title{Adaptive Fuzzy Tracking Control with Global Prescribed-Time Prescribed Performance for Uncertain Strict-Feedback Nonlinear Systems}

\author{Bing Mao, Xiaoqun Wu, Hui Liu,~\IEEEmembership{Member,~IEEE,} Yuhua Xu, Jinhu L\"{u},~\IEEEmembership{Fellow,~IEEE,}
\thanks{Bing Mao is with the School of Mathematics
and Statistics, Wuhan University, Hubei 430072, China (e-mail:
bmaomath@whu.edu.cn).}
\thanks{Xiaoqun Wu is with the School of Mathematics and Statistics, Wuhan University, Hubei 430072, China, with Hubei Key Laboratory of Computational Science, Wuhan University, Hubei 430072, China, and also with Research Center of Complex Network, Wuhan University, Hubei 430072, China (e-mail: xqwu@whu.edu.cn).}
\thanks{Hui Liu is with the School of Automation,
Huazhong University of Science and Technology, Hubei 430074, China,
and also with the Key Laboratory of Image Processing and Intelligent
Control of Education Ministry of China, Hubei 430074, China (e-mail:
hliu@hust.edu.cn).}
\thanks{Yuhua Xu is with the School of Statistics and Data Science, Nanjing Audit University, Jiangsu 211815, China (e-mail: yuhuaxu2004@163.com)}
\thanks{Jinhu L\"{u} is with the School of Automation Science and Electrical
Engineering, Beihang University, Beijing 100191, China (e-mail:
jhlu@iss.ac.cn).}}

\markboth{T\MakeLowercase {his work has been submitted to the} IEEE \MakeLowercase {for possible publication.} C\MakeLowercase {opyright may be transferred without notice, after which this version may no longer be accessible.}}%
{}


\maketitle

\begin{abstract}
  Adaptive fuzzy control strategies are established to achieve global prescribed performance with prescribed-time convergence for strict-feedback systems with mismatched uncertainties and unknown nonlinearities. Firstly, to quantify the transient and steady performance constraints of the tracking error, a class of prescribed-time prescribed performance functions are designed, and a novel error transformation function is introduced to  remove the initial value constraints and solve the singularity problem in existing works. Secondly, based on dynamic surface control methods, controllers with or without approximating structures are established to guarantee that the tracking error achieves prescribed transient performance and converges into a prescribed bounded set within prescribed time. In particular, the settling time and initial value of the prescribed performance function are completely independent of initial conditions of the tracking error and system parameters, which improves existing results.  Moreover, with a novel Lyapunov-like energy function, not only the differential explosion problem frequently occurring in backstepping techniques is solved, but the drawback of the semi-global boundedness of tracking error induced by dynamic surface control can be overcome. The validity and effectiveness of the main results are verified by numerical simulations on practical examples.
\end{abstract}


\section{Introduction}
\label{sec1}
\IEEEPARstart{T}{rajectory} tracking is one of the most fundamental problems in control community with wide applications in many fields, including leader-following tracking \cite{Wu2022Mao,Salmanpour2022MehdiArefi}, formation-containment tracking \cite{Zhang2022Wang,Lu2022Dong,Liu2021Wu}, bipartite tracking \cite{Liang2021Guo,Liu2022Basin}, average tracking \cite{Sen2022Sahoo,Zhao2022Xian} and synchronization \cite{Xu2022WuMao,Xu2022WuLi,Li2021Wu}. To guarantee the tracking performance, prescribed performance control \cite{Bechlioulis2008Rovithakis} has been commonly employed, which can ensure the convergence of the tracking error into a sufficiently small prescribed residual set with a prescribed convergence rate and a desired transient state performance. In the past decades, many efforts have been devoted to prescribed performance control and fruitful results have been developed.

A robust adaptive controller with exponential prescribed performance function for unknown multi-input multi-output (MIMO) systems was developed in 2008 \cite{Bechlioulis2008Rovithakis}, which pioneered the methodology of prescribed performance control. Subsequently, the prescribed performance control for MIMO nonlinear systems \cite{Bechlioulis2010Rovithakis,Kostarigka2012Rovithakis,Theodorakopoulos2015Rovithakis,Shi2021,Gao2022Li}, uncertain strict-feedback systems with unknown control directions \cite{Zhang2017Yang,Li2018Liu,Zhao2022Song,Zhang2021Wang,Zhang2019Wang}, and high-power nonlinear systems \cite{Lv2022Chen}  were studied using different control techniques. With hysteretic actuator
nonlinearity and faults, the problem of adaptive fuzzy prescribed performance control for nonlinear systems was solved by using the command filter theory \cite{Zhang2018Yang}. In \cite{Wang2016Wang}, the prescribed performance control was extended to the leader-following consensus for uncertain nonlinear strict-feedback multiagent systems under directed communication graphs. Nussbaum-type functions and fuzzy logic systems were introduced to solve the problem of unknown control directions and nonlinearities, respectively. However, the Nussbaum-type function expends the dynamic order of the closed-loop systems and fuzzy logic systems lead to semi-global boundedness of all closed-loop signals. To this end, decentralized control laws of low complexity in the sense of no prior knowledge of system nonlinearity, no approximating structures, no complex calculations and static control protocols, were proposed \cite{Bechlioulis2017Rovithakis,Katsoukis2022Rovithakis}.

The event-triggered control, as an effective energy-saving scheme, was introduced to explore adaptive fuzzy prescribed performance control strategies for pure-feedback nonlinear systems with unknown nonlinearities and unmeasured states \cite{Qiu2019Sun}. In \cite{Dong2020Gao}, fuzzy adaptive dynamic surface control was employed to solve the differential explosion problem of backstepping techniques \cite{Zhao2022Song,Qiu2019Sun}, and an error-driven nonlinear feedback function was designed to establish the semi-global stability of closed-loop systems. Via uniting control \cite{Kanakis2020Rovithakis}, the global prescribed performance of the output tracking error was achieved, improving the semi-global stability \cite{Kostarigka2012Rovithakis,Theodorakopoulos2015Rovithakis,Shi2021,Gao2022Li,Wang2016Wang,Zhang2018Yang,Qiu2019Sun,Dong2020Gao,Li2020Shao}. Moreover, significant modifications on the standard prescribed performance control methodology were provided to successfully handle discontinuities in desired trajectory \cite{Fotiadis2021Rovithakis}, and time-varying delays in both state measurements and control inputs \cite{Bikas2021Rovithakis}.

It should be emphasized that only asymptotic or exponential convergence of the tracking error system can be guaranteed in all aforementioned results, which restricts their practical applicability under finite time constraints. Accordingly, a finite-time performance function is defined in \cite{Liu2018Liu}, and semi-globally practical finite-time tracking of a class of uncertain non-strict feedback nonlinear systems is achieved using an adaptive neural network controller. Following this idea, the finite-time prescribed performance control is studied for non-strict feedback nonlinear systems with adaptive fuzzy control \cite{Liu2019Liu,Sui2022Tong}, multivariable strict feedback nonlinear systems with neural network control \cite{Jiang2022Huang}, strict feedback nonlinear systems with disturbance observer-based control \cite{Qiu2022Wang}, and stochastic nonlinear systems with adaptive backstepping control \cite{Sui2021Chen}. Based on exponential performance functions, finite-time adaptive fuzzy control with event-triggered mechanism for uncertain strict-feedback nonlinear systems \cite{Sun2021Qiu} and multiagent systems \cite{Zhou2022Sui} and a fixed-time version for uncertain robot systems \cite{Zhu2022Yang} are established. Particularly, fuzzy logic systems \cite{Liu2019Liu} and dynamic surface control \cite{Sui2022Tong,Jiang2022Huang,Qiu2022Wang,Sun2021Qiu,Zhou2022Sui} are used to deal with the differential explosion problem, and thus complex calculations can be avoided.

Finite-/fixed-time stability ensures higher convergence accuracy, faster convergence rate and better anti-interference ability than asymptotic or exponential ones, but the settling-time function is neither independent of the initial states nor the system parameters. Therefore, the settling time generally cannot be prescribed in advance by users. It is even nontrivial to obtain the explicit convergence time under unavailable initial states or system parameters.

At present, few results have been reported about the prescribed-time prescribed performance control. Using a skillful rate function and a self-tuning Nussbaum-type function, the problem is preliminarily solved for a second-order Euler-Lagrange system with full-state constraints and nonparametric uncertainties \cite{Zhao2018Song}. Then, this methodology is extended to uncertain strict-feedback nonlinear systems \cite{Zhao2019Song,Shao2022Wang,Cao2022Cao} and high-order multi-agent systems \cite{Cao2021Song} via traditional backstepping control techniques. However, the self-tuning Nussbaum-type function multiplies the dynamic order of the closed-loop systems, which, together with the differential explosion problem, leads to higher computational complexity. Besides, the time-varying performance functions introduced in \cite{Shao2022Wang,Cao2022Cao} are singular at the initial time, which further limits its practical applicability. In conclusion, how to obtain globally prescribed-time prescribed performance without high computational complexity for uncertain strick-feedback nonlinear systems is still an open problem and deserves further investigation.

Inspired by aforementioned discussions, this work focuses on designing an adaptive fuzzy tracking control as well as reduce computational complexity to achieve global prescribed-time prescribed performance for strict-feedback nonlinear systems with unknown nonlinearities and mismatched uncertainties. In existing work, there are three open issues that may result in semi-global boundedness of closed-loop signals.
\begin{itemize}
  \item[(A1)] To well define the error transformation function at the initial time, the initial value of the prescribed performance function is related to one of the tracking error \cite{Kostarigka2012Rovithakis,Theodorakopoulos2015Rovithakis,Wang2016Wang,Bechlioulis2017Rovithakis,Katsoukis2022Rovithakis,Gao2022Li,
      Shi2021,Zhang2018Yang,Qiu2019Sun,Kanakis2020Rovithakis,Fotiadis2021Rovithakis,Bikas2021Rovithakis,Sui2021Chen,Qiu2022Wang,
      Sun2021Qiu,Zhou2022Sui,Zhu2022Yang,Lv2022Chen,Cao2021Song,Zhang2019Wang}.
  \item[(A2)] To suppress the influence of the error surface defined in dynamic surface control on the energy function, derivatives of virtual controllers are bounded or defined on a compact set \cite{Liu2019Liu,Shojaei2019Arefi,Qiu2022Wang}.
  \item[(A3)] The system states are constrained in a compact set while using neural networks or fuzzy logic systems to approximate the unknown nonlinearities \cite{Kostarigka2012Rovithakis,Theodorakopoulos2015Rovithakis,Wang2016Wang,Shi2021,Zhang2018Yang,Dong2020Gao,Qiu2019Sun,Li2020Shao,
      Liu2018Liu,Liu2019Liu,Sui2022Tong,Jiang2022Huang,Sui2021Chen,Qiu2022Wang,Sun2021Qiu,Zhou2022Sui,Zhu2022Yang}.
\end{itemize}
Therefore, in addition to avoiding high computational complexity caused by the differential explosion and Nussbaum-type functions, it is also theoretically challenging to guarantee global performance. The main contributions of this paper and comparisons with some related works are summarized as follows.
\begin{itemize}
  \item[1)] A novel prescribed-time prescribed performance function is defined. Contrary to finite-time prescribed functions \cite{Liu2018Liu,Liu2019Liu,Sui2022Tong,Jiang2022Huang,Qiu2022Wang,Sui2021Chen}, the settling time is independent of initial conditions and system parameters and can be prescribed in advance by users. In addition, the problem in (A1) is solved due to the design of a novel error transformation function.
  \item[2)] A novel Lyapunov-like energy function is proposed to avoid the differential explosion. A skillful time-varying function from the derivative of the Lyapunov-like energy function is used to eliminate the influence of the error surface, which solves the problem in (A2). Besides, the problem in (A3) is addressed using a generalized Lipschitz condition.
  \item[3)] Two fuzzy control strategies with or without approximating structures are established, achieving global prescribed performance of tracking error, and guaranteeing the global uniform boundedness of all closed-loop signals. Specifically, no Nussbaum-type functions are used and no singular phenomenon occurs in control design. Consequently, the proposed controllers are superior to those in \cite{Zhao2018Song,Zhao2019Song,Shao2022Wang,Cao2021Song,Cao2022Cao} in terms of reducing computational complexity and improving practical implementability.
\end{itemize}

The remainder of this work is organized as follows. Preliminaries are presented in Section \ref{sec2}. Control design and stability analysis are provided in Section \ref{sec3}, and two practical examples are presented to verify the validity and effectiveness of the proposed methods in Section \ref{sec4}.  Section \ref{sec5} concludes this paper.

\emph{Notation:} Let $R_{\geq0}$ and $R^n$ denote the set of non-negative real numbers and the $n-$dimensional Euclidean space, respectively. $I_m$ is an $m-$ dimensional identity matrix, and $\mathbf{1}_n $ $(\mathbf{0}_n)$ stands for a vector with all entries equal to $1$ $(0)$.

\section{Preliminaries and Problem Formulation}
\label{sec2}
In this section, the system model, fuzzy logic systems, some lemmas and assumptions are provided.
\subsection{System Descriptions}
\label{sec2-1}
Consider a strict-feedback nonlinear system with mismatched uncertainties
\begin{align}
\label{equ1}
\dot{x}_i(t)&=f_i(\bar{x}_i(t))+g_i(\bar{x}_i(t))x_{i+1}(t)+\omega_i(t),1\leq i\leq n-1,\notag\\
\dot{x}_n(t)&=f_n(\bar{x}_n(t))+g_n(\bar{x}_n(t))u(t)+\omega_n(t),\\
y(t)&=x_1(t)\notag,
\end{align}
where $x_i(t)\in R$ is the system state, $\bar{x}_i(t)=(x_1(t),\cdots,x_i(t))^\top\in R^i$, $u(t)\in R$, $\omega_i(t)\in R$ and $y(t)\in R$ denote the control input, external disturbance and output trajectory of the system, respectively. $f_i(\bar{x}_i(t)): R^i\rightarrow R$ and $g_i(\bar{x}_i(t)): R^i\rightarrow R$ are unknown continuous nonlinear functions, called nonlinearities and control coefficients, respectively.

Define the reference signal as $y_r(t)$, which is bounded, continuous and differentiable, and denote $\bar{y}_i(t)=(y_r(t),\cdots,y_r(t))^\top\in R^i$.

\theoremseparator{\it{:}}
\newcounter{remark1}
\addtocounter{remark1}{0}
\newtheorem*{remark1}{\it Remark 1}
\begin{remark1}
\label{rem1}
\textup{
Compared with \cite{Zhang2017Yang,Li2018Liu,Zhao2022Song,Qiu2019Sun,Kanakis2020Rovithakis,Zhao2018Song,Qiu2022Wang,Shao2022Wang,Liu2019Liu}, system \eqref{equ1} is more common and can describe many practical control plants including robot manipulators, mass-spring-damper systems, parallel active suspension systems, ship maneuvering systems and switched RLC circuits. Therefore, it is of great significance to study the tracking control problem  of system \eqref{equ1}, specifically, with prescribed-time prescribed performance for more desired system response.
}
\end{remark1}
\subsection{Fuzzy Logic Systems}
\label{sec2-2}
A fuzzy logic system is composed of fuzzifier, fuzzy rule base, fuzzy inference engine and defuzzifier based on the fuzzy if-then rules \cite{Wang1992Mendel}:
\par
$Q_j:$ if $\varpi_1$ is $\Omega_1^j$, $\varpi_2$ is $\Omega_2^j$, $\cdots$, and $\varpi_n$ is $\Omega_n^j$,
\par\qquad
then $\varsigma$ is $\Delta^j$, $j=1, 2,\cdots, m$,
\\
where $\varpi=(\varpi_1, \varpi_2, \cdots, \varpi_n)^\top\in R^n$, $\varsigma\in R$ and $m$ are the input, output variables and the number of fuzzy rules, respectively, $\Omega_i^j$ and $\Delta^j$ $(i=1, \cdots, n, j=1,2, \cdots, m)$ are the fuzzy sets with fuzzy membership functions $\mu_{\Omega_i^j}(\varpi_i)$ and $\mu_{\Delta^j}(\varsigma)$, respectively. In general, the approximation of a fuzzy logic system can be described as via center average defuzzification,  product inference, singleton fuzzifiers and Gaussian membership functions:
\begin{align*}
\varsigma(\varpi)=\phi^\top(\varpi)\theta=\sum_{j=1}^r\vartheta_j\phi_j(\varpi),
\end{align*}
where $\phi(\varpi)=(\phi_1(\varpi), \cdots, \phi_m(\varpi))^\top$, $\theta=(\vartheta_1, \cdots, \vartheta_m)^\top$, and $\vartheta_j$ $(j=1, 2,\cdots, m)$ are optimal weights and
\begin{align*}
\phi_j(\varpi)=\frac{\Pi_{i=1}^n\mu_{\Omega_i^j}(\varpi_i)}{\sum_{j=1}^m\left(\Pi_{i=1}^n\mu_{\Omega_i^j}(\varpi_i)\right)}
~(j=1,2, \cdots, m)
\end{align*}
are fuzzy basis functions. Singleton fuzzifiers and Gaussian membership functions are defined by
\begin{align*}
\mu_{\Delta^j}(\varsigma)=
\begin{cases}
1, \text{ if } \varsigma=\vartheta_j\\
0, \text{ if } \varsigma\neq \vartheta_j
\end{cases} {\rm and~}
\mu_{\Omega_i^j}(\varpi_i)=a_i^j {\rm e}^{-\frac{1}{2}\left(\frac{\varpi_i-\bar{\varpi}_i^j}{\sigma_i^j}\right)^2},
\end{align*}
respectively, where $a_i^j>0$, $\bar{\varpi}_i^j$ and $\sigma_i^j$  $(i=1, 2,\cdots, n, j=1, 2,\cdots, m)$ are some constants.

\subsection{Assumptions and Lemmas}
\label{sec2-3}
Before proceeding further, some common assumptions and lemmas are necessary.
\theoremseparator{\it{:}}
\newcounter{assumption1}
\addtocounter{assumption1}{0}
\newtheorem*{assumption1}{\it Assumption 1}
\begin{assumption1}
\label{assum1}
\textup{
The reference signal $y_r(t)$ and $\dot{y}_r(t)$ are bounded and available for control design.
}
\end{assumption1}

\theoremseparator{\it{:}}
\newcounter{assumption2}
\addtocounter{assumption2}{1}
\newtheorem*{assumption2}{\it Assumption 2}
\begin{assumption2}
\label{assum2}
\textup{
The sign of function $g_i(\bar{x}_i(t))$ is certain definite. Without loss of generality, suppose that there exist positive constants $\underline{g_i}$ and $\bar{g}_i$ such that $0<\underline{g_i}<g_i(\bar{x}_i(t))<\bar{g}_i<+\infty$.
}
\end{assumption2}

\theoremseparator{\it{:}}
\newcounter{assumption3}
\addtocounter{assumption3}{2}
\newtheorem*{assumption3}{\it Assumption 3}
\begin{assumption3}
\label{assum3}
\textup{
There exists an unknown constant $\bar{\omega}_i>0$ such that $\lvert \omega_i(t)\rvert \leq \bar{\omega}_i$ $(i= 1, \cdots,n)$.
}
\end{assumption3}

\theoremseparator{\it{:}}
\newcounter{assumption4}
\addtocounter{assumption4}{3}
\newtheorem*{assumption4}{\it Assumption 4}
\begin{assumption4}
\label{assum4}
\textup{
There exists a positive and continuous function $L_i(x_i(t),y_i(t),t)$ such that for any $x_i\in R^i$ and $y_i\in R^i$,
\begin{align*}
\lvert f_i(x_i(t))-f_i(y_i(t))\rvert\leq L_i(x_i(t),y_i(t),t)\lVert x_i(t)-y_i(t)\rVert
\end{align*}
holds $(i=1,\cdots,n)$, where $L_i(x_i(t),y_i(t),t)$ is bounded if $x_i(t)$ and $y_i(t)$ are bounded.
}
\end{assumption4}

\theoremseparator{\it{:}}
\newcounter{remark2}
\addtocounter{remark2}{1}
\newtheorem*{remark2}{\it Remark 2}
\begin{remark2}
\label{rem2}
\textup{
Assumption \ref{assum1} implies that for any $\bar{y}_i\in R^i$ and $t\in [0,+\infty)$, there must exist some compact set $\Omega_i\subset R^i$ such that $\bar{y}_i\in \Omega_i$, which makes it feasible to introduce fuzzy logic systems to deal with the unknown nonlinearity. Assumptions \ref{assum2} and \ref{assum3} are used in most existing results \cite{Liu2019Liu,Dong2020Gao,Fotiadis2021Rovithakis,Cui2020Xia,Cai2022Shi}. Assumption \ref{assum4} can be seen as a generalized Lipschitz condition which is less restrictive as compared with that in \cite{Zhao2019Song,Shao2022Wang}.
}
\end{remark2}

\theoremseparator{\it{(\cite{Krstic1995Kanellakopoulos}):}}
\newcounter{lemma1}
\addtocounter{lemma1}{0}
\newtheorem*{lemma1}{\it Lemma 1}
\begin{lemma1}
\label{lem1}
\textup{Let $\epsilon>0$, $a>1$, $b>1$ satisfying $(a-1)(b-1)=1$. Then, for all $x,y\in R$,
\begin{align*}
xy\leq \frac{\epsilon^a}{a}\lvert x\lvert^a+\frac{1}{b\epsilon^b}\lvert y\lvert^b.
\end{align*}
}
\end{lemma1}

\theoremseparator{\it{(\cite{Ma2020Ma}):}}
\newcounter{lemma2}
\addtocounter{lemma2}{1}
\newtheorem*{lemma2}{\it Lemma 2}
\begin{lemma2}
\label{lem2}
\textup{For any constants $\alpha>0$ and $\beta\in R$,
\begin{align*}
0\leq \lvert \beta\rvert -\frac{\beta^2}{\sqrt{\beta^2+\alpha^2}}\leq \alpha.
\end{align*}
}
\end{lemma2}

\theoremseparator{\it{(\cite{Wang1992Mendel,Deng2018Yang}):}}
\newcounter{lemma3}
\addtocounter{lemma3}{2}
\newtheorem*{lemma3}{\it Lemma 3}
\begin{lemma3}
\label{lem3}
\textup{Consider a continuous nonlinear function $f(x(t)): \Omega\rightarrow R$ defined on a compact set $\Omega\subset R^n$. Then, there exists a fuzzy logic system with bounded optimal weight $\theta=(\vartheta_1, \cdots, \vartheta_m)^\top$ and fuzzy basis function $\phi(x(t))=(\phi_1(x(t)), \cdots, \phi_m(x(t)))^\top$ such that for any $\bar{\varepsilon}>0$,
\begin{align*}
f(x(t))=\phi^\top(x(t))\theta+\varepsilon(t),
\end{align*}
where $ \varepsilon(t)$ is the approximation error satisfying $\lvert \varepsilon(t)\lvert\leq \bar{\varepsilon}$ for all $t\in[0,+\infty)$. In addition, for $\phi(x(t))$, there holds
\begin{align*}
\phi(x(t))\phi^\top(x(t))\leq m I_m.
\end{align*}
}
\end{lemma3}

\theoremseparator{\it{(\cite{Ge2004Wang}):}}
\newcounter{lemma4}
\addtocounter{lemma4}{3}
\newtheorem*{lemma4}{\it Lemma 4}
\begin{lemma4}
\label{lem4}
\textup{For continuous function $V(t): R_{\geq0}\rightarrow R_{\geq0}$ and bounded function $\varsigma(t):R_{\geq0}\rightarrow R$, if there exist constants $d_1>0$ and $d_2\in R$ such that
\begin{align*}
\dot{V}(t)\leq -d_1V(t)+d_2\varsigma(t),
\end{align*}
then $V(t)$ is bounded.
}
\end{lemma4}

\section{Main Results}
\label{sec3}
This section presents main results including prescribed performance function, control scheme design and stability analysis.
\subsection{Prescribed Performance Functions}
\label{sec3-1}
The subsection introduces a novel prescribed-time performance function and further discusses some useful properties.

\theoremseparator{\it{:}}
\newcounter{definition1}
\addtocounter{definition1}{0}
\newtheorem*{definition1}{\it Definition 1}
\begin{definition1}
\label{def1}
\textup{
Let $\underline{\eta}>0$ be a positive constant and $T>0$ be any user-prescribed time. A continuous and differentiable function $\eta(t)$ is called the prescribed-time performance function if satisfying
\begin{itemize}
\item[(i)] $\eta(t)>0$ for all $t\in [0,+\infty)$;
\item[(ii)] $\dot{\eta}(t)\leq 0$;
\item[(iii)] $\lim_{t\rightarrow T}\eta(t)=\underline{\eta}$ and $\eta(t)\equiv \underline{\eta}$ for all $t\geq T$,
\end{itemize}
where $T$ and $\underline{\eta}$ stand for prescribed time and prescribed accuracy, respectively.
}
\end{definition1}

\theoremseparator{\it{:}}
\newcounter{remark3}
\addtocounter{remark3}{2}
\newtheorem*{remark3}{\it Remark 3}
\begin{remark3}
\label{rem3}
\textup{
Compared with the finite-time performance function \cite{Liu2019Liu,Sui2021Chen,Qiu2022Wang}, the settling time $T$ of the prescribed-time performance function is independent of initial values and system parameters, which contributes to simplifying the design of performance function, improving convergence rate and achieving global performance. Therefore, it can better meet the practical application demands.
}
\end{remark3}

From Definition \ref{def1}, a typical prescribed-time performance function can be designed as
\begin{align}
\label{equ2}
\eta(t)=
\begin{cases}
a{\rm e}^{-b\big(\frac{T}{T-t}\big)^h}+c,&0\leq t<T,\\
c,&t\geq T,
\end{cases}
\end{align}
where $h>0$, $a,b$ and $c$ are positive constants satisfying $\eta(0)=a{\rm e}^{-b}+c=\frac{\pi}{2}$.

Define the tracking error $e(t)=y(t)-y_r(t)$ and the error transformation function
\begin{align}
\label{equ3}
z_1(t)=\tan\bigg(\frac{\pi}{2}\frac{\arctan(e(t))}{\eta(t)}\bigg),
\end{align}
which means that
\begin{align}
\label{equ4}
e(t)&=\tan\bigg(\frac{2}{\pi}\eta(t)\arctan(z_1(t))\bigg).
\end{align}
Therefore,
\begin{align}
\label{equ5}
\dot{e}(t)&=\frac{\frac{2}{\pi}\dot{\eta}(t)\arctan(z_1(t))+\frac{2}{\pi}\eta(t)\frac{\dot{z}_1(t)}{1+z_1^2(t)}}{\cos^2\big(\frac{2\eta(t)\arctan(z_1(t))}{\pi}\big)},
\end{align}
which is equivalent to
\begin{align}
\label{equ6}
\dot{z}_1(t)=&\psi(t)\big(\dot{e}(t)\varphi(t)-\frac{2}{\pi}\dot{\eta}(t)\arctan(z_1(t))\big)\notag\\
=&\psi(t)\big((f_1(\bar{x}_1(t))+g_1(\bar{x}_1(t))x_{2}+\omega_1(t)\notag\\
&-\dot{y}_r(t))\varphi(t)-\frac{2}{\pi}\dot{\eta}(t)\arctan(z_1(t))\big),
\end{align}
where
\begin{align*}
\psi(t)=\frac{\pi(1+z_1^2(t))}{2\eta(t)}>0
\end{align*}
and
\begin{align*}
\varphi(t)=\cos^2\bigg(\frac{2\eta(t)\arctan(z_1(t))}{\pi}\bigg)>0.
\end{align*}
\theoremseparator{\it{:}}
\newcounter{proposition1}
\addtocounter{proposition1}{0}
\newtheorem*{proposition1}{\it Proposition 1}
\begin{proposition1}
\label{prop1}
\textup{
The following properties hold for functions $\eta(t)$ and $z_1(t)$:
\begin{itemize}
  \item[(i)] $\eta(t)$ is continuous and infinitely differentiable with
  \begin{align*}
  \dot{\eta}(t)=
  \begin{cases}
  -\frac{abhT^h}{(T-t)^{h+1}}{\rm e}^{-b\big(\frac{T}{T-t}\big)^h},&0\leq t<T,\\
  0&t\geq T;
  \end{cases}
  \end{align*}
  \item[(ii)]  $z_1(0)=e(0)$ for any $e(0)\in R$, and $z(t)$ is well defined if $\lvert e(t)\rvert<\tan(\eta(t))$;
  \item[(iii)] if $z_1(t)$ is bounded for all $t\in [0,+\infty)$, then one has transient state performance
  $\lvert e(t)\rvert<\tan(\eta(t))$ for all $t\geq 0$, and steady state performance $\lvert e(t)\rvert<\tan(c)$ for all $t\geq T$.
\end{itemize}
}
\end{proposition1}

\theoremseparator{\it{:}}
\newcounter{remark4}
\addtocounter{remark4}{3}
\newtheorem*{remark4}{\it Remark 4}
\begin{remark4}
\label{rem4}
\textup{
It is worth noting that in the existing exponential \cite{Zhang2019Wang,Zhang2017Yang,Wang2016Wang,Lv2022Chen,Bechlioulis2017Rovithakis,Katsoukis2022Rovithakis,Gao2022Li,Shi2021,Zhang2018Yang,Qiu2019Sun,Kanakis2020Rovithakis } and finite-time \cite{Sui2021Chen,Qiu2022Wang,Cao2021Song} prescribed performance control, the error transformation depends on the initial values of tracking error $e(t)$ and performance function $\eta(t)$, i.e., $\lvert e(0)\rvert <\eta(0)$, which may lead to semi-global stability of the tracking error and impose some difficulties on the implementation of the error transformation in the absence of initial values. Prescribed-time performance functions in \cite{Shao2022Wang,Cao2022Cao} are infinite at the initial time in order to guarantee $\lvert e(0)\rvert <\eta(0)$, which leads to the singularity problem. The error transformation \eqref{equ3} addresses the above problems via tangent function $\tan(\cdot)$ and its inverse. Another common barrier function is hyperbolic arctangent function ${\rm arctanh}(\cdot)$. It should be pointed out that only symmetrical prescribed performance can be obtained via tangent function and hyperbolic arctangent function, which may not be applicable for some specific scenarios. Two kinds of asymmetrical prescribed performance can be achieved via combining the tangent function and the hyperbolic arctangent function, which are
\begin{align*}
(i)~\bar{z}_1(t)&=
\begin{cases}
\tan\big(\frac{\pi}{2}\frac{\arctan(e(t))}{\eta(t)}\big),~e(t)\geq 0,\\
{\rm arctanh}\big(\frac{2}{\pi}\frac{\tanh(e(t))}{\eta(t)}\big),~e(t)<0,
\end{cases}\text{and}\\
(ii)~\hat{z}_1(t)&=\begin{cases}
{\rm arctanh}\big(\frac{2}{\pi}\frac{\tanh(e(t))}{\eta(t)}\big),~e(t)\geq 0,\\
\tan\big(\frac{\pi}{2}\frac{\arctan(e(t))}{\eta(t)}\big),~e(t)< 0.
\end{cases}
\end{align*}
It can be observed that both $\bar{z}_1(t)$ and $\hat{z}_1(t)$ are continuously differentiable.
}
\end{remark4}

According to Proposition \ref{prop1}, the control objective is given as follows.
\theoremseparator{\it{:}}
\newcounter{objective1}
\addtocounter{objective1}{0}
\newtheorem*{objective1}{\it Objective 1}
\begin{objective1}
\label{obje1}
\textup{
Design controller $u(t)$ to guarantee that
\begin{itemize}
  \item[(i)] the tracking error $e(t)$ converges to a prescribed region with a prescribed time, and satisfies transient state performance $\lvert e(t)\rvert<\tan(\eta(t))$ for all $t\geq 0$ and steady state performance $\lvert e(t)\rvert<\tan(c)$ for all $t\geq T$.
  \item[(ii)] all signals in system \eqref{equ1} are globally and uniformly bounded.
\end{itemize}
}
\end{objective1}
\subsection{Control Schemes}
\label{sec3-2}
In the subsection, to avoid the differential explosion problem in traditional backstepping techniques, a novel dynamic surface control is developed to design adaptive fuzzy controller to guarantee the tracking error with global prescribed-time prescribed performance.

Let $\alpha_{i-1}(t)$ be the virtual control, and define the intermediate error $z_i(t)=x_i(t)-s_i(t)$ and the error surface $r_i(t)=s_i(t)-\alpha_{i-1}(t)~(i=2,3,\cdots,n)$, where $s_i$ is the filtering signal obtained by the first-order filter $\lambda_i\dot{s}_i(t)+s_i(t)=\alpha_{i-1}(t)$ with positive constant $\lambda_i>0$ and initial condition $s_i(0)=\alpha_{i-1}(0)$. Then, the control design is divided into the following three steps.

\emph{Step 1:}
According to Eq. \eqref{equ6}, the transformed closed-loop system is
\begin{align}
\label{equ7}
\dot{z}_1(t)=&\psi(t)\bigg(\big(f_1(\bar{x}_1(t))-f_1(\bar{y}_1(t))+f_1(\bar{y}_1(t))\notag\\
&+g_1(\bar{x}_1(t))x_{2}+\omega_1(t)-\dot{y}_r(t)\big)\varphi(t)\notag\\
&-\frac{2}{\pi}\dot{\eta}(t)\arctan(z(t))\bigg).
\end{align}
It follows from Assumption \ref{assum1} and Lemma \ref{lem3} that for any given estimate accuracy $\bar{\varepsilon}_1>0$, there exist a optimal weight $\theta_1\in R^m$ and a fuzzy basis function $\phi_1(\bar{y}_1(t)): R\rightarrow R^m$ such that
\begin{align}
\label{equ8}
f_1(\bar{y}_1(t))=\phi_1^\top(\bar{y}_1(t))\theta_1+\varepsilon_1(t),
\end{align}
where $\lvert \varepsilon_1(t)\rvert\leq \bar{\varepsilon}_1$. Let $\hat{\theta}_1(t)\in R^m$ be the estimate of the optimal weight $\theta_1$ and $\tilde{\theta}_1(t)=\theta_1-\hat{\theta}_1(t)$ be the estimate error. Consider the Lyapunov-like  energy function
\begin{align}
\label{equ9}
V_1(t)=&\frac{1}{2}z_1^2(t)+\frac{1}{2\mu_1}\tilde{\theta}_1^\top(t) \tilde{\theta}_1(t),
\end{align}
where $\mu_1>0$ is a positive constant. Differentiating $V_1(t)$ along the solution of closed-loop system \eqref{equ7} yields
\begin{align}
\label{equ10}
\dot{V}_1(t)=&z_1(t)\psi(t)\bigg(\big(f_1(\bar{x}_1(t))-f_1(\bar{y}_1(t))
+\phi_1^\top(\bar{y}_1(t))\tilde{\theta}_1(t)\notag\\
&+\phi_1^\top(\bar{y}_1(t))\hat{\theta}_1(t)
+\varepsilon_1(t)+g_1(\bar{x}_1(t))x_2(t)\notag\\
&+\omega_1(t)-\dot{y}_r(t)\big)\varphi(t)-\frac{2}{\pi}\dot{\eta}(t)\arctan(z_1(t))\bigg)\notag\\
&-\frac{1}{\mu_1}\tilde{\theta}_1^\top(t)\dot{\hat{\theta}}_1(t)\notag\\
=&z_1(t)\psi(t)\bigg(\big(f_1(\bar{x}_1(t))-f_1(\bar{y}_1(t))+\phi_1^\top(\bar{y}_1(t))\tilde{\theta}_1(t)\notag\\
&+\phi_1^\top(\bar{y}_1(t))\hat{\theta}_1(t)+\varepsilon_1(t)\notag\\
&+g_1(\bar{x}_1(t))(z_2(t)+r_2(t)+\alpha_1(t))\notag\\
&+\omega_1(t)-\dot{y}_r(t)\big)\varphi(t)-\frac{2}{\pi}\dot{\eta}(t)\arctan(z_1(t))\bigg)\notag\\
&
-\frac{1}{\mu_1}\tilde{\theta}_1^\top(t) \dot{\hat{\theta}}_1(t)\notag\\
\leq&\lvert z_1(t)\rvert\varphi(t)\psi(t)\lvert f_1(\bar{x}_1(t))-f_1(\bar{y}_1(t))\rvert\notag\\
&+z_1(t)\varphi(t)\psi(t)\phi_1^\top(\bar{y}_1(t))\tilde{\theta}_1(t)
+\big(z_1(t)\varphi(t)\psi(t)\big)^2\notag\\
&+z_1(t)\psi(t)\big((\phi_1^\top(\bar{y}_1(t))\hat{\theta}_1(t)-\dot{y}_r(t))\varphi(t)\notag\\
&-\frac{2}{\pi}\dot{\eta}(t)\arctan(z_1(t))\big)-\frac{1}{\mu_1}\tilde{\theta}_1^\top(t) \dot{\hat{\theta}}_1(t)\notag\\
&+\lvert z_1(t)\rvert\varphi(t)\psi(t)\bar{g}_1(\lvert z_2(t)\rvert+\lvert r_2(t)\rvert)\notag\\
&+\frac{1}{2}(\bar{\varepsilon}_1^2+\bar{\omega}_1^2)+z_1(t)\varphi(t)\psi(t)g_1(\bar{x}_1(t))\alpha_1(t),
\end{align}
where Assumption \ref{assum3} and Lemma \ref{lem1} are used to derive the last inequality. Design
\begin{align}
\label{equ11}
\alpha_1(t)=&-\frac{z_1(t)\varphi(t)\psi(t)\beta_1^2(t)}{\underline{g_1}\sqrt{(z_1(t)\varphi(t)\psi(t)\beta_1(t))^2+\delta_1^2}}\notag\\
&-\frac{z_1(t)\varphi(t)\psi(t)\chi_1^2(t)}{\underline{g_1}\sqrt{(z_1(t)\varphi(t)\psi(t)\chi_1(t))^2+\sigma_1^2}}\notag\\
&-\frac{z_1(t)\varphi(t)\psi(t)}{\underline{g_1}}
-\frac{\varpi_1z_1(t)}{2\underline{g_1}\varphi(t)\psi(t)},\\
\label{equ12}
\dot{\hat{\theta}}_1(t)=&-\varpi_1\hat{\theta}_1(t)
+\mu_1z_1(t)\varphi(t)\psi(t)\phi_1(\bar{y}_1(t)),
\end{align}
where $\varpi_1, \delta_1$ and $\sigma_1$ are some positive constants, $\beta_1(t)$ and $\chi_1(t)$ will be designed later. Therefore, the last term of Eq. \eqref{equ10} can be calculated as
\begin{align}
\label{equ13}
&z_1(t)\varphi(t)\psi(t)g_1(\bar{x}_1(t))\alpha_1(t)\notag\\
=&-\frac{g_1(\bar{x}_1(t))(z_1(t)\varphi(t)\psi(t)\beta_1(t))^2}{\underline{g_1}\sqrt{(z_1(t)\varphi(t)\psi(t)\beta_1(t))^2+\delta_1^2}}\notag\\
&-\frac{g_1(\bar{x}_1(t))(z_1(t)\varphi(t)\psi(t)\chi_1(t))^2}{\underline{g_1}\sqrt{(z_1(t)\varphi(t)\psi(t)\chi_1(t))^2+\sigma_1^2}}\notag\\
&-\frac{g_1(\bar{x}_1(t))}{\underline{g_1}}\big(z_1(t)\varphi(t)\psi(t)\big)^2\notag\\
&-\frac{\varpi_1g_1(\bar{x}_1(t))z_1^2(t)}{2\underline{g_1}}\notag\\
\leq&-\frac{(z_1(t)\varphi(t)\psi(t)\beta_1(t))^2}{\sqrt{(z_1(t)\varphi(t)\psi(t)\beta_1(t))^2+\delta_1^2}}\notag\\
&-\frac{(z_1(t)\varphi(t)\psi(t)\chi_1(t))^2}{\sqrt{(z_1(t)\varphi(t)\psi(t)\chi_1(t))^2+\sigma_1^2}}\notag\\
&-\big(z_1(t)\varphi(t)\psi(t)\big)^2-\frac{\varpi_1z_1^2(t)}{2}\notag\\
\leq&\delta_1-\lvert z_1(t)\varphi(t)\psi(t)\beta_1(t)\rvert\notag\\
&+\sigma_1-\lvert z_1(t)\varphi(t)\psi(t)\chi_1(t)\rvert\notag\\
&-\big(z_1(t)\varphi(t)\psi(t)\big)^2-\frac{\varpi_1z_1^2(t)}{2}\notag\\
\leq&- z_1(t)\varphi(t)\psi(t)\beta_1(t)+\delta_1\notag\\
&-\lvert z_1(t)\varphi(t)\psi(t)\chi_1(t)\rvert+\sigma_1\notag\\
&-\big(z_1(t)\varphi(t)\psi(t)\big)^2-\frac{\varpi_1z_1^2(t)}{2},
\end{align}
where Assumption \ref{assum2}, Lemmas \ref{lem2} and \ref{lem1} are employed to obtain the first, second and last inequalities, respectively. Consequently, combining Eqs. \eqref{equ10}, \eqref{equ12}  and \eqref{equ13} yields
\begin{align}
\label{equ14}
\dot{V}_1(t)\leq&\lvert z_1(t)\rvert\varphi(t)\psi(t)(\lvert f_1(\bar{x}_1(t))-f_1(\bar{y}_1(t))\rvert -\lvert\chi_1(t)\rvert) \notag\\
&+\lvert z_1(t)\rvert\varphi(t)\psi(t)\bar{g}_1(\lvert z_2(t)\rvert+\lvert r_2(t)\rvert)\notag\\
&+z_1(t)\psi(t)\big((\phi_1^\top(\bar{y}_1(t))\hat{\theta}_1(t)-\beta_1(t)-\dot{y}_r(t))\varphi(t)\notag\\
&-\frac{2}{\pi}\dot{\eta}(t)\arctan(z_1(t))\big)-\frac{\varpi_1z_1^2(t)}{2}\notag\\
&\frac{\varpi_1}{\mu_1}\tilde{\theta}_1^\top(t) \hat{\theta}_1(t)+\frac{1}{2}(\bar{\varepsilon}_1^2+\bar{\omega}_1^2)+\delta_1+\sigma_1.
\end{align}
Define
\begin{align}
\label{equ15}
\beta_1(t)=&\phi_1^\top(\bar{y}_1(t))\hat{\theta}_1(t)-\dot{y}_r(t)\notag\\
&-\frac{2}{\pi\varphi(t)}\dot{\eta}(t)\arctan(z_1(t)),\\
\label{equ16}
\chi_1(t)=&L_1(\bar{x}_1(t),\bar{y}_1(t),t)\lVert \bar{x}_1(t)-\bar{y}_1(t)\rVert.
\end{align}
From Assumption \ref{assum4}, Eqs. \eqref{equ15} and \eqref{equ16}, one has
\begin{align}
\label{equ17}
\dot{V}_1(t)\leq&\lvert z_1(t)\rvert\varphi(t)\psi(t)\bar{g}_1(\lvert z_2(t)\rvert+\lvert r_2(t)\rvert)\notag\\
&-\frac{\varpi_1z_1^2(t)}{2}+\frac{1}{2}(\bar{\varepsilon}_1^2+\bar{\omega}_1^2)+\delta_1+\sigma_1\notag\\
&+\frac{\varpi_1}{\mu_1}\tilde{\theta}_1^\top(t)\hat{\theta}_1(t)\notag\\
=&\lvert z_1(t)\rvert\varphi(t)\psi(t)\bar{g}_1(\lvert z_2(t)\rvert+\lvert r_2(t)\rvert)-\frac{\varpi_1z_1^2(t)}{2}\notag\\
&+\frac{1}{2}(\bar{\varepsilon}_1^2+\bar{\omega}_1^2)+\delta_1+\sigma_1+\frac{\varpi_1}{2\mu_1}\theta_1^\top\theta_1\notag\\
&-\frac{\varpi_1}{2\mu_1}\hat{\theta}_1^\top(t)\hat{\theta}_1(t)-\frac{\varpi_1}{2\mu_1}\tilde{\theta}_1^\top(t)\tilde{\theta}_1(t)\notag\\
\leq&-\varpi_1V_1(t)+\Lambda_1\notag\\
&+\lvert z_1(t)\rvert\varphi(t)\psi(t)\bar{g}_1(\lvert z_2(t)\rvert+\lvert r_2(t)\rvert),
\end{align}
where $\Lambda_1=\frac{1}{2}(\bar{\varepsilon}_1^2+\bar{\omega}_1^2)+\delta_1+\sigma_1+\frac{\varpi_1}{2\mu_1}\theta_1^\top\theta_1$.

\emph{Step 2:} For $i=2,3,\cdots,n-1$, according to $z_i(t)=x_i(t)-s_i(t)$, one has
\begin{align}
\label{equ18}
\dot{z}_i(t)=&\dot{x}_i(t)-\dot{s}_i(t)\notag\\
=&f_i(\bar{x}_i(t))+g_i(\bar{x}_i(t))x_{i+1}+\omega_i(t)-\dot{s}_i(t)\notag\\
=&f_i(\bar{x}_i(t))+g_i(\bar{x}_i(t))(z_{i+1}(t)+r_{i+1}(t)+\alpha_i(t))\notag\\
&+\omega_i(t)-\dot{s}_i(t)\notag\\
=&f_i(\bar{x}_i(t))-f_i(\bar{y}_i(t))+\phi_i^\top(\bar{y}_i(t))\theta_i+\varepsilon_i(t)\notag\\
&+g_i(\bar{x}_i(t))(z_{i+1}(t)+r_{i+1}(t)+\alpha_i(t))\notag\\
&+\omega_i(t)-\dot{s}_i(t),
\end{align}
where $f_i(\bar{y}_i(t))=\phi_i^\top(\bar{y}_i(t))\theta_i+\varepsilon_i(t)$ is employed with $\lvert\varepsilon_i(t)\rvert\leq \bar{\varepsilon}_i$ under Assumption \ref{assum1} and Lemma \ref{lem3}. Let $\hat{\theta}_i(t)\in R^m$ be the estimate of the optimal weight $\theta_i$ and $\tilde{\theta}_i(t)=\theta_i-\hat{\theta}_i(t)$ be the estimate error. Consider the Lyapunov-like energy function
\begin{align}
\label{equ19}
V_i(t)=V_{i-1}(t)+{\rm arctan}(z_i(t))+\varrho_i\lvert z_i(t)\rvert+\frac{1}{2\mu_i}\tilde{\theta}_i^\top(t)\tilde{\theta}_i(t),
\end{align}
where $\mu_i>0$ and $\varrho_i>1$ are positive constants. Denote $\zeta_i(t)=\frac{1}{1+z_i^2(t)}+\varrho_i{\rm sign}(z_i(t))$. Then, $\zeta_i(t)\neq0$ for all $z_i(t)$. Differentiating $V_i(t)$ along the solution of closed-loop system \eqref{equ20} yields
\begin{align}
\label{equ20}
\dot{V}_i(t)=&\dot{V}_{i-1}(t)+\zeta_i(t)\dot{z}_i(t)-\frac{1}{\mu_i}\tilde{\theta}_i^\top(t)\dot{\hat{\theta}}_i(t)\notag\\
=&\dot{V}_{i-1}(t)+\zeta_i(t)\big(f_i(\bar{x}_i(t))-f_i(\bar{y}_i(t))\notag\\
&+\phi_i^\top(\bar{y}_i(t))\tilde{\theta}_i(t)+\phi_i^\top(\bar{y}_i(t))\hat{\theta}_i(t)\notag\\
&+g_i(\bar{x}_i(t))(z_{i+1}(t)+r_{i+1}(t)+\alpha_i(t))\notag\\
&+\varepsilon_i(t)+\omega_i(t)-\dot{s}_i(t)\big)-\frac{1}{\mu_i}\tilde{\theta}_i^\top(t)\dot{\hat{\theta}}_i(t)\notag\\
\leq&\dot{V}_{i-1}(t)+\lvert \zeta_i(t)(f_i(\bar{x}_i(t))-f_i(\bar{y}_i(t)))\rvert \notag\\
&+\zeta_i(t)\big(\phi_i^\top(\bar{y}_i(t))\tilde{\theta}_i(t)+\phi_i^\top(\bar{y}_i(t))\hat{\theta}_i(t)\notag\\
&-\frac{1}{\lambda_i}(\alpha_{i-1}(t)-s_i(t))\big)-\frac{1}{\mu_i}\tilde{\theta}_i^\top(t)\dot{\hat{\theta}}_i(t)\notag\\
&+\bar{g}_i\lvert \zeta_i(t)\rvert(\lvert z_{i+1}(t)\rvert+\lvert r_{i+1}(t)\rvert)\notag\\
&+\zeta_i(t)g_i(\bar{x}_i(t))\alpha_i(t)+\zeta_i^2(t)+\frac{1}{2}(\bar{\varepsilon}_i^2+\bar{\omega}_i^2),
\end{align}
where the last inequality is obtained via Lemma \ref{lem1} and Assumption \ref{assum2}.

Design
\begin{align}
\label{equ21}
\alpha_i(t)=&-\frac{\zeta_i(t)\beta_i^2(t)}{\underline{g_i}\sqrt{\zeta_i^2(t)\beta_i^2(t)+\delta_i^2}}
-\frac{\zeta_i(t)\chi_i^2(t)}{\underline{g_i}\sqrt{\zeta_i^2(t)\chi_i^2(t)+\sigma_i^2}}\notag\\
&-\frac{\zeta_i(t)\gamma_i^2(t)}{\underline{g_i}\sqrt{\zeta_i^2(t)\gamma_i^2(t)+\rho_i^2}}
-\frac{\zeta_i(t)\xi_i^2(t)}{\underline{g_i}\sqrt{\zeta_i^2(t)\gamma_i^2(t)+\tau_i^2}}\notag\\
&-\frac{\varpi_i\big({\rm arctan}(z_i(t))+\varrho_i\lvert z_i(t)\rvert\big)}{\underline{g_i}\zeta_i(t)}-\frac{\zeta_i(t)}{\underline{g_i}},\\
\label{equ22}
\dot{\hat{\theta}}_i(t)=&-\varpi_i\hat{\theta}_i(t)+\mu_i\zeta_i(t)\phi_i(\bar{y}_i(t)),
\end{align}
where $\varpi_i,$ $\delta_i$, $\sigma_i$, $\rho_i$ and $\tau_i$ are some positive constants, and
\begin{align}
\label{equ23}
\beta_i(t)=&\phi_i^\top(\bar{y}_i(t))\hat{\theta}_i(t)-\frac{1}{\lambda_i}(\alpha_{i-1}(t)-s_i(t)),\\
\label{equ24}
\gamma_i(t)=&\begin{cases}
\frac{\bar{g}_1\varphi(t)\psi(t)\lvert z_1(t)z_2(t)\rvert}{\zeta_2(t)},i=2,\\
\frac{\bar{g}_{i-1}\lvert \zeta_{i-1}(t) z_i(t)\rvert}{\zeta_i(t)},i>2,
\end{cases}\\
\label{equ25}
\chi_i(t)=&L_i(\bar{x}_i(t),\bar{y}_i(t),t)\lVert \bar{x}_i(t)-\bar{y}_i(t)\rVert,\\
\label{equ26}
\xi_i(t)=&\begin{cases}
\frac{\bar{g}_1\varphi(t)\psi(t)\lvert z_1(t)r_2(t)\rvert}{\zeta_2(t)},i=2,\\
\frac{\bar{g}_{i-1}\lvert \zeta_{i-1}(t)r_i(t)\rvert}{\zeta_i(t)},i>2.
\end{cases}
\end{align}
According to Eq. \eqref{equ21}, the sixth term of Eq. \eqref{equ20} can be calculated as
\begin{align}
\label{equ27}
&\zeta_i(t)g_i(\bar{x}_i(t))\alpha_i(t)\notag\\
=&-\frac{g_i(\bar{x}_i(t))\zeta_i^2(t)\beta_i^2(t)}{\underline{g_i}\sqrt{\zeta_i^2(t)\beta_i^2(t)+\delta_i^2}}
-\frac{g_i(\bar{x}_i(t))\zeta_i^2(t)\chi_i^2(t)}{\underline{g_i}\sqrt{\zeta_i^2(t)\chi_i^2(t)+\sigma_i^2}}\notag\\
&-\frac{g_i(\bar{x}_i(t))\zeta_i^2(t)\gamma_i^2(t)}{\underline{g_i}\sqrt{\zeta_i^2(t)\gamma_i^2(t)+\rho_i^2}}
-\frac{g_i(\bar{x}_i(t))\zeta_i^2(t)\xi_i^2(t)}{\underline{g_i}\sqrt{\zeta_i^2(t)\gamma_i^2(t)+\tau_i^2}}\notag\\
&-\frac{\varpi_ig_i(\bar{x}_i(t))\big({\rm arctan}(z_i(t))+\varrho_i\lvert z_i(t)\rvert\big)}{\underline{g_i}}\notag\\
&-\frac{g_i(\bar{x}_i(t))\zeta_i^2(t)}{\underline{g_i}}\notag\\
\leq&-\frac{\zeta_i^2(t)\beta_i^2(t)}{\sqrt{\zeta_i^2(t)\beta_i^2(t)+\delta_i^2}}
-\frac{\zeta_i^2(t)\chi_i^2(t)}{\sqrt{\zeta_i^2(t)\chi_i^2(t)+\sigma_i^2}}\notag\\
&-\frac{\zeta_i^2(t)\gamma_i^2(t)}{\sqrt{\zeta_i^2(t)\gamma_i^2(t)+\rho_i^2}}
-\frac{\zeta_i^2(t)\xi_i^2(t)}{\sqrt{\zeta_i^2(t)\gamma_i^2(t)+\tau_i^2}}\notag\\
&-\varpi_i\big({\rm arctan}(z_i(t))+\varrho_i\lvert z_i(t)\rvert\big)-\zeta_i^2(t)\notag\\
\leq&\delta_i-\lvert \zeta_i(t)\beta_i(t)\rvert+\sigma_i-\lvert \zeta_i(t)\chi_i(t)\rvert\notag\\
&+\rho_i-\lvert \zeta_i(t)\gamma_i(t)\rvert+\tau_i-\lvert \zeta_i(t)\xi_i(t)\rvert\notag\\
&-\varpi_i\big({\rm arctan}(z_i(t))+\varrho_i\lvert z_i(t)\rvert\big)-\zeta_i^2(t)\notag\\
\leq&-\zeta_i(t)\beta_i(t)-\lvert \zeta_i(t)\chi_i(t)\rvert-\lvert \zeta_i(t)\gamma_i(t)\rvert\notag\\
&-\lvert \zeta_i(t)\xi_i(t)\rvert-\varpi_i\big({\rm arctan}(z_i(t))+\varrho_i\lvert z_i(t)\rvert\big)\notag\\
&-\zeta_i^2(t)+\delta_i+\sigma_i+\rho_i+\tau_i,
\end{align}
where Assumption \ref{assum2} and Lemma \ref{lem2} are used to derive the first and second inequality, respectively. Therefore, according to Assumption \ref{assum3}, combining Eqs. \eqref{equ20}, \eqref{equ21}, \eqref{equ22} and \eqref{equ27} yields
\begin{align}
\label{equ28}
\dot{V}_i(t)\leq&\dot{V}_{i-1}(t)-\varpi_i\big({\rm arctan}(z_i(t))+\varrho_i\lvert z_i(t)\rvert\big)\notag\\
&+\frac{\varpi_i}{\mu_i}\tilde{\theta}_i^\top(t)\hat{\theta}_i(t)+\bar{g}_i\lvert \zeta_i(t)\rvert(\lvert z_{i+1}(t)\rvert+\lvert r_{i+1}(t)\rvert)\notag\\
&-\lvert \zeta_i(t)\gamma_i(t)\rvert-\lvert \zeta_i(t)\xi_i(t)\rvert+\frac{1}{2}(\bar{\varepsilon}_i^2+\bar{\omega}_i^2)\notag\\
&+\delta_i+\sigma_i+\rho_i+\tau_i\notag\\
=&-\sum_{k=2}^{i-1}\varpi_k\bigg({\rm arctan}(z_k(t))+\varrho_k\lvert z_k(t)\rvert\notag\\
&+\frac{\varpi_k}{2\mu_k}\hat{\theta}_k^\top(t)\hat{\theta}_k(t)\bigg)-\varpi_1V_1(t)+\sum_{k=1}^{i-1}\Lambda_k\notag\\
&-\varpi_i\big({\rm e}^{z_i(t)}+\varrho_i\lvert z_i(t)\rvert\big)+\frac{\varpi_i}{2\mu_i}\theta_i^\top\theta_i\notag\\
&-\frac{\varpi_i}{2\mu_i}\hat{\theta}_i^\top(t)\hat{\theta}_i(t)
-\frac{\varpi_i}{2\mu_i}\tilde{\theta}_i^\top(t)\tilde{\theta}_i(t)\notag\\
&+\bar{g}_i\lvert \zeta_i(t)\rvert(\lvert z_{i+1}(t)\rvert+\lvert r_{i+1}(t)\rvert)\notag\\
&+\frac{1}{2}(\bar{\varepsilon}_i^2+\bar{\omega}_i^2)+\delta_i+\sigma_i+\rho_i+\tau_i\notag\\
\leq&-\sum_{k=2}^i\varpi_k\bigg({\rm arctan}(z_k(t))+\varrho_k\lvert z_k(t)\rvert\notag\\
&+\frac{\varpi_k}{2\mu_k}\hat{\theta}_k^\top(t)\hat{\theta}_k(t)\bigg)-\varpi_1V_1(t)+\sum_{k=1}^i\Lambda_k\notag\\
&+\bar{g}_i\lvert \zeta_i(t)\rvert(\lvert z_{i+1}(t)\rvert+\lvert r_{i+1}(t)\rvert),
\end{align}
where $\Lambda_k=\frac{1}{2}(\bar{\varepsilon}_k^2+\bar{\omega}_k^2)+\delta_k+\sigma_k+\rho_k+\tau_k+\frac{\varpi_k}{2\mu_k}\theta_k^\top\theta_k~(2\leq k\leq n-1)$.

\emph{Step 3:} For $i=n$, $z_n(t)=x_n(t)-s_n(t)$,
\begin{align}
\label{equ29}
\dot{z}_n(t)=&\dot{x}_n(t)-\dot{s}_n(t)\notag\\
=&f_n(\bar{x}_n(t))+g_n(\bar{x}_n(t))u(t)+\omega(t)-\dot{s}_n(t)\notag\\
=&f_n(\bar{x}_n(t))-f_n(\bar{y}_n(t))+\phi_n^\top(\bar{y}_n(t))\theta_n+\varepsilon_n(t)\notag\\
&+g_n(\bar{x}_n(t))u(t)+\omega_n(t)-\dot{s}_n(t),
\end{align}
where $f_n(\bar{y}_n(t))=\phi_n^\top(\bar{y}_n(t))\theta_n+\varepsilon_n(t)$ is employed with $\lvert\varepsilon_n(t)\rvert\leq \bar{\varepsilon}_n$ under Assumption \ref{assum1} and Lemma \ref{lem3}. Let $\hat{\theta}_n(t)\in R^m$ be the estimate of the optimal weight $\theta_n$ and $\tilde{\theta}_n(t)=\theta_n-\hat{\theta}_n(t)$ be the estimate error. Consider the energy function
\begin{align}
\label{equ30}
V_n(t)=V_{n-1}(t)+{\rm arctan}(z_n(t))+\varrho_n\lvert z_n(t)\rvert+\frac{1}{2\mu_n}\tilde{\theta}_n^\top(t)\tilde{\theta}_n(t),
\end{align}
where $\mu_n>0$ and $\varrho_n>1$ are positive constants. Denote $\zeta_n(t)=\frac{1}{1+z_n^2(t)}+\varrho_n{\rm sign}(z_n(t))$. Then, $\zeta_n(t)\neq0$ for all $z_n(t)$. Differentiating $V_n(t)$ along the solution of closed-loop system \eqref{equ29} yields
\begin{align}
\label{equ31}
\dot{V}_n(t)=&\dot{V}_{n-1}(t)+\zeta_n(t)\big(f_n(\bar{x}_n(t))-f_n(\bar{y}_n(t))\notag\\
&+\phi_n^\top(\bar{y}_n(t))\tilde{\theta}_n(t)+\phi_n^\top(\bar{y}_n(t))\hat{\theta}_n(t)+\varepsilon_n(t)\notag\\
&+g_n(\bar{x}_n(t))u(t)+\omega_n(t)-\dot{s}_n(t)\big)-\frac{1}{\mu_n}\tilde{\theta}_n^\top(t)\dot{\hat{\theta}}_n(t)\notag\\
\leq&\dot{V}_{n-1}(t)+\lvert \zeta_n(t)(f_n(\bar{x}_n(t))-f_n(\bar{y}_n(t)))\rvert \notag\\
&+\zeta_n(t)\big(\phi_n^\top(\bar{y}_n(t))\tilde{\theta}_n(t)+\phi_n^\top(\bar{y}_n(t))\hat{\theta}_n(t)\notag\\
&-\frac{1}{\lambda_n}(\alpha_{n-1}(t)-s_n(t))\big)-\frac{1}{\mu_n}\tilde{\theta}_n^\top(t)\dot{\hat{\theta}}_n(t)\notag\\
&+\zeta_n(t)g_n(\bar{x}_n(t))u(t)+\zeta_n^2(t)+\frac{1}{2}(\bar{\varepsilon}_n^2+\bar{\omega}_n^2).
\end{align}
Design the actual controllers as follows
\begin{align}
\label{equ32}
u(t)=&-\frac{\zeta_n(t)\beta_n^2(t)}{\underline{g_n}\sqrt{\zeta_n^2(t)\beta_n^2(t)+\delta_n^2}}
-\frac{\zeta_n(t)\chi_n^2(t)}{\underline{g_n}\sqrt{\zeta_n^2(t)\chi_n^2(t)+\sigma_n^2}}\notag\\
&-\frac{\zeta_n(t)\gamma_n^2(t)}{\underline{g_n}\sqrt{\zeta_n^2(t)\gamma_n^2(t)+\rho_n^2}}
-\frac{\zeta_n(t)\xi_n^2(t)}{\underline{g_n}\sqrt{\zeta_n^2(t)\gamma_i^2(t)+\tau_n^2}}\notag\\
&-\frac{\varpi_n\big({\rm arctan}(z_n(t))+\varrho_n\lvert z_n(t)\rvert\big)}{\underline{g_n}\zeta_n(t)}-\frac{\zeta_n(t)}{\underline{g_n}},\\
\label{equ33}
\dot{\hat{\theta}}_n(t)=&-\varpi_n\hat{\theta}_n(t)+\mu_n\zeta_n(t)\phi_n(\bar{y}_n(t)),
\end{align}
where $\varpi_n,$ $\delta_n$, $\sigma_n$, $\rho_n$ and $\tau_n$ are some positive constants, and
\begin{align}
\label{equ34}
\beta_n(t)=&\phi_n^\top(\bar{y}_n(t))\hat{\theta}_n(t)-\frac{1}{\lambda_n}(\alpha_{n-1}(t)-s_n(t)),\\
\label{equ35}
\gamma_n(t)=&
\frac{\bar{g}_{n-1}\lvert \zeta_{n-1}(t)z_n(t)\rvert}{\zeta_n(t)},\\
\label{equ36}
\chi_n(t)=&L_n(\bar{x}_n(t),\bar{y}_n(t),t)\lVert \bar{x}_n(t)-\bar{y}_n(t)\rVert,\\
\label{equ37}
\xi_n(t)=&
\frac{\bar{g}_{n-1}\lvert \zeta_{n-1}(t)r_n(t)\rvert}{\zeta_n(t)}.
\end{align}
Similar to Eq. \eqref{equ27}, one has
\begin{align}
\label{equ38}
\zeta_n(t)g_n(\bar{x}_n(t))u(t)
\leq&-\zeta_n(t)\beta_n(t)-\lvert \zeta_n(t)\chi_n(t)\rvert\notag\\
&-\lvert \zeta_n(t)\gamma_n(t)\rvert-\lvert \zeta_n(t)\xi_n(t)\rvert\notag\\
&-\varpi_n\big({\rm arctan}(z_n(t))+\lvert z_n(t)\rvert\big)\notag\\
&-\zeta_n^2(t)+\delta_n+\sigma_n+\rho_n+\tau_n.
\end{align}
Therefore, it follows from Eqs. \eqref{equ31}, \eqref{equ32}, \eqref{equ33} and \eqref{equ38} that
\begin{align}
\label{equ39}
\dot{V}_n(t)\leq&\dot{V}_{n-1}(t)-\lvert \zeta_n(t)\gamma_n(t)\rvert-\lvert \zeta_n(t)\xi_n(t)\rvert \notag\\
&-\varpi_n\big({\rm e}^{z_n(t)}+\lvert z_n(t)\rvert\big)+\frac{1}{\mu_n}\tilde{\theta}_n^\top(t)\hat{\theta}_n(t)\notag\\
&+\frac{1}{2}(\bar{\varepsilon}_n^2+\bar{\omega}_n^2)+\delta_n+\sigma_n+\rho_n+\tau_n\notag\\
\leq&-\sum_{k=2}^n\varpi_k\bigg({\rm arctan}(z_k(t))+\varrho_n\lvert z_k(t)\rvert\notag\\
&+\frac{\varpi_k}{2\mu_k}\hat{\theta}_k^\top(t)\hat{\theta}_k(t)\bigg)-\varpi_1V_1(t)+\sum_{k=1}^n\Lambda_k\notag\\
\leq&-\varpi V_n(t)+\Lambda,
\end{align}
where $\varpi=\min\{\varpi_1,\cdots,\varpi_n\}$, $\Lambda_n=\frac{1}{2}(\bar{\varepsilon}_n^2+\bar{\omega}_n^2)+\delta_n+\sigma_n+\rho_n+\tau_n$ and $\Lambda=\sum_{k=1}^n\Lambda_k$.

The main results are presented in the following theorems.

\theoremseparator{\it{:}}
\newcounter{theorem1}
\addtocounter{theorem1}{0}
\newtheorem*{theorem1}{\it Theorem 1}
\begin{theorem1}
\label{theo1}
\textup{
Suppose that Assumptions \ref{assum1}, \ref{assum2}, \ref{assum3} and \ref{assum4} hold. Using the virtual controllers \eqref{equ11} and \eqref{equ21}, adaptive fuzzy update laws \eqref{equ12}, \eqref{equ22} and \eqref{equ33}, the actual controller \eqref{equ32} achieves Objective \ref{obje1}.
}
\end{theorem1}
\begin{IEEEproof}
Consider the energy function \eqref{equ30}. Then, it follows from Eq. \eqref{equ39} and Lemma \ref{lem4} that $V_n(t)\leq V_n(0){\rm e}^{-\varpi t}+\frac{\Lambda}{\varpi}\big(1-{\rm e}^{-\varpi t}\big)$. Therefore, both $z_i(t)$ and $\tilde{\theta}_i(t)$ $(i=1,2,\cdots,n)$ are uniformly ultimately bounded. According to Proposition \ref{prop1}, Objective \ref{obje1}(i) is achieved.

From $\tilde{\theta}_1(t)=\theta_1-\hat{\theta}_1(t)$, the boundedness of $\tilde{\theta}_1(t)$ implies the boundedness of $\hat{\theta}_1(t)$. Therefore, it follows from  Lemma \ref{lem3} that
\begin{align}
\label{equ40}
\big(\phi_1^\top(\bar{y}_1(t))\hat{\theta}_1(t)\big)^2=&\big(\phi_1^\top(\bar{y}_1(t))\hat{\theta}_1(t)\big)^\top\big(\phi_1^\top(\bar{y}_1(t))\hat{\theta}_1(t)\big)\notag\\
=&\hat{\theta}_1^\top(t)\big(\phi_1(\bar{y}_1(t))\phi_1^\top(\bar{y}_1(t))\big)\hat{\theta}_1(t)\notag\\
\leq&m\hat{\theta}_1^\top(t)\hat{\theta}_1(t)<+\infty,
\end{align}
which guarantees the boundedness of $\beta_1(t)$. Furthermore, from $e(t)=x_1(t)-y_r(t)$ and Assumption \ref{assum1}, $\chi_i(t)$ in Eq. \eqref{equ16} is bounded. Therefore, the virtual control $\alpha_1(t)$ is bounded based on Eqs. \eqref{equ15} and \eqref{equ16}, which implies that
\begin{align*}
s_2(t)=&\alpha_1(0){\rm e}^{-\frac{t}{\lambda_2}}+\int_0^t\alpha_1(\tau){\rm e}^{-\frac{t-\tau}{\lambda_2}}d\tau\\
\leq&\alpha_1(0){\rm e}^{-\frac{t}{\lambda_2}}+\hat{\alpha}_1\big(1-{\rm e}^{-\frac{t}{\lambda_2}}\big),
\end{align*}
where $\hat{\alpha}_1=\sup_{t\in[0,+\infty)}\{\lvert \alpha_1(t)\rvert\}$. Consequently, the filtering signal $s_2(t)$ is bounded. Recursively, the boundedness of $\beta_i(t)$, $\chi_i(t)$, $\xi_i(t)$, $r_i(t)$, $x_i(t)$ and $s_i(t)$ $(i=2,3,\cdots,n)$ can be guaranteed via $z_i(t)=x_i(t)-s_i(t)$ and $ r_i(t)=s_i(t)-\alpha_{i-1}(t)$, and no finite-time escape phenomenon can occur. The proof is thus completed.
\end{IEEEproof}

Generally, employing adaptive fuzzy estimator to acquire the information of unknown nonlinearity may increase the computational complexity, which will cause unnecessary consumption of computational power and equipment wear. To avoid this, let
\begin{align}
\label{equ41}
\beta_1(t)=&z_1(t)\varphi(t)\psi(t)\phi_1^\top(\bar{y}_1(t))\phi_1(\bar{y}_1(t))\notag\\
&-\dot{y}_r(t)-\frac{2}{\pi\varphi(t)}\dot{\eta}(t)\arctan(z_1(t)),\\
\label{equ42}
\beta_i(t)=&\zeta_i(t)\phi_i^\top(\bar{y}_i(t))\phi_i(\bar{y}_i(t))\notag\\
&-\frac{1}{\lambda_i}(\alpha_{i-1}(t)-s_i(t)),1< i<n,\\
\label{equ43}
\beta_n(t)=&\zeta_n(t)\phi_n^\top(\bar{y}_n(t))\phi_n(\bar{y}_n(t))\notag\\
&-\frac{1}{\lambda_n}(\alpha_{n-1}(t)-s_n(t)).
\end{align}
Then, the results with prescribed-time prescribed performance without approximating structures can be developed.

\theoremseparator{\it{:}}
\newcounter{theorem2}
\addtocounter{theorem2}{1}
\newtheorem*{theorem2}{\it Theorem 2}
\begin{theorem2}
\label{theo2}
\textup{
Suppose that Assumptions \ref{assum1}, \ref{assum2}, \ref{assum3} and \ref{assum4} hold. Under the virtual controllers \eqref{equ11} and \eqref{equ21} with $\beta_1(t)$ and $\beta_i(t)$ defined in Eqs. \eqref{equ41} and \eqref{equ42}, respectively, the conclusions in Theorem \ref{theo1} hold via the actual controller \eqref{equ32} with $\beta_n(t)$ in Eq. \eqref{equ43}.
}
\end{theorem2}
\begin{IEEEproof}
Consider the energy function
\begin{align}
\label{equ44}
V(t)=\sum_{i=1}^nV_i(t),
\end{align}
where $V_1(t)=\frac{1}{2}z_1^2(t)$ and $V_i(t)={\rm arctan}(z_i(t))+\varrho_i\lvert z_i(t)\rvert~(i=2,3,\cdots,n)$. Differentiating $V_1(t)$ along system \eqref{equ7} yields
\begin{align}
\label{equ45}
\dot{V}_1(t)=&z_1(t)\psi(t)\bigg(\big(f_1(\bar{x}_1(t))-f_1(\bar{y}_1(t))\notag\\
&+\phi_1^\top(\bar{y}_1(t))\theta_1+\varepsilon_1(t)+\omega_1(t)\notag\\
&+g_1(\bar{x}_1(t))(z_2(t)+r_2(t)+\alpha_1(t))\notag\\
&-\dot{y}_r(t)\big)\varphi(t)-\frac{2}{\pi}\dot{\eta}(t)\arctan(z_1(t))\bigg)\notag\\
\leq&\lvert z_1(t)\varphi(t)\psi(t)(f_1(\bar{x}_1(t))-f_1(\bar{y}_1(t)))\rvert\notag\\
&+z_1^2(t)\psi^2(t)\varphi^2(t)\big(1+\frac{1}{2}\phi_1^\top(\bar{y}_1(t))\phi_1(\bar{y}_1(t))\big)\notag\\
&+\frac{1}{2}(\theta_1^\top\theta_1+\bar{\varepsilon}_1^2+\bar{\omega}_1^2)\notag\\
&+\bar{g}_1\varphi(t)\psi(t)\lvert z_1(t)\rvert(\lvert z_2(t)\rvert+\lvert r_2(t)\rvert)\notag\\
&+z_1(t)\varphi(t)\psi(t)g_1(\bar{x}_1(t))(\alpha_1(t)-\dot{y}_r(t))\notag\\
&-\frac{2}{\pi}z_1(t)\psi(t)\dot{\eta}(t)\arctan(z_1(t))\notag\\
\leq&-\varpi_1 V_1(t)+\Delta_1,
\end{align}
where $\Delta_1=\frac{1}{2}(\bar{\varepsilon}_1^2+\bar{\omega}_1^2)+\delta_1+\sigma_1+\frac{1}{2}\theta_1^\top\theta_1$.

Similar to the analytical derivation of Eq. \eqref{equ39}, one has
\begin{align}
\label{equ46}
\dot{V}(t)\leq-\varpi V(t)+\Delta,
\end{align}
where $\Delta=\sum_{k=1}^{n}\Delta_k$ and $\Delta_k=\delta_k+\sigma_k+\rho_j+\tau_k+\frac{1}{2}\bar{\varepsilon}_k^2+\frac{1}{2}\bar{\omega}_k^2+\frac{1}{2}\theta_k^\top\theta_k~ (2\leq k\leq n)$. The proof is thus completed
\end{IEEEproof}

\theoremseparator{\it{:}}
\newcounter{remark5}
\addtocounter{remark5}{4}
\newtheorem*{remark5}{\it Remark 5}
\begin{remark5}
\label{rem5}
\textup{
In traditional dynamic surface control \cite{Qiu2022Wang,Shojaei2019Arefi}, the error surface $r_i(t)$ plays a significant part in the boundeness of the energy function. This is because $r_i(t)$  is an integral part of the energy function, which means that it has got to guarantee the boundedness of $\dot{\alpha}_i(t)$, but this is typically challenging and may cause semiglobal stability of closed-loop systems. In this work, a novel function, $\zeta_i(t)=\frac{1}{1+z_i^2(t)}+\varrho_i{\rm sign}(z_i(t))$, is exploited to eliminate the effect of the error surface on the energy function, which makes it feasible to achieve global stability of closed-loop systems when solving the differential explosion problem of backstepping techniques.
}
\end{remark5}

\theoremseparator{\it{:}}
\newcounter{remark6}
\addtocounter{remark6}{5}
\newtheorem*{remark6}{\it Remark 6}
\begin{remark6}
\label{rem6}
\textup{
There are many tunable parameters in prescribed performance function and control schemes including $a,b,c,h$ and $T$ in performance function, $\delta_i, \sigma_i, \rho_i, \tau_i, \varrho_i, \mu_i$ and $\varpi_i$ in controllers and  adaptive fuzzy update laws, and $\lambda_i$ in the first-order filter. Positive constants $a,b,c,h$ and $T$ are preassigned by users according to practical control demands under the constraint $a{\rm e}^{-b} +c=\frac{\pi}{2}$. From Proposition \ref{prop1}, $\tan(c)$ is the maximum convergence threshold of the tracking error, and it thus should select $c$ according to practical requirements. Parameters $\mu_i$ and $\varpi_i$ have an impact on the update rate of the adaptive fuzzy estimate $\hat{\theta}_i(t)$, but do not require careful adjustment as long as $\mu_i>0$ and $\varpi_i>0$. The steady state performance has nothing to do with the values of $\delta_i, \sigma_i, \rho_i$ and $\tau_i$, which means that these parameters can be chosen as some positive constants big enough to reduce the values of virtual and actual controllers. Besides, $\lim_{z_i(t)\rightarrow 0^-}\zeta_i(t)=1-\varrho_i\neq0$, and equivalently, $\lim_{z_i(t)\rightarrow 0^-}\frac{1}{\zeta_i(t)} =\frac{1}{1-\varrho_i}\neq \infty$, which explains why $\varrho_i>0$ and $\varrho_i\neq 1$.
}
\end{remark6}

\section{Simulatons}
\label{sec4}
In this section, some examples are provided to demonstrate the validity and performance of the proposed methods.

\begin{figure}
  \centering
  \includegraphics[width=8cm,height=4.5cm,trim=150 50 30 20,clip]{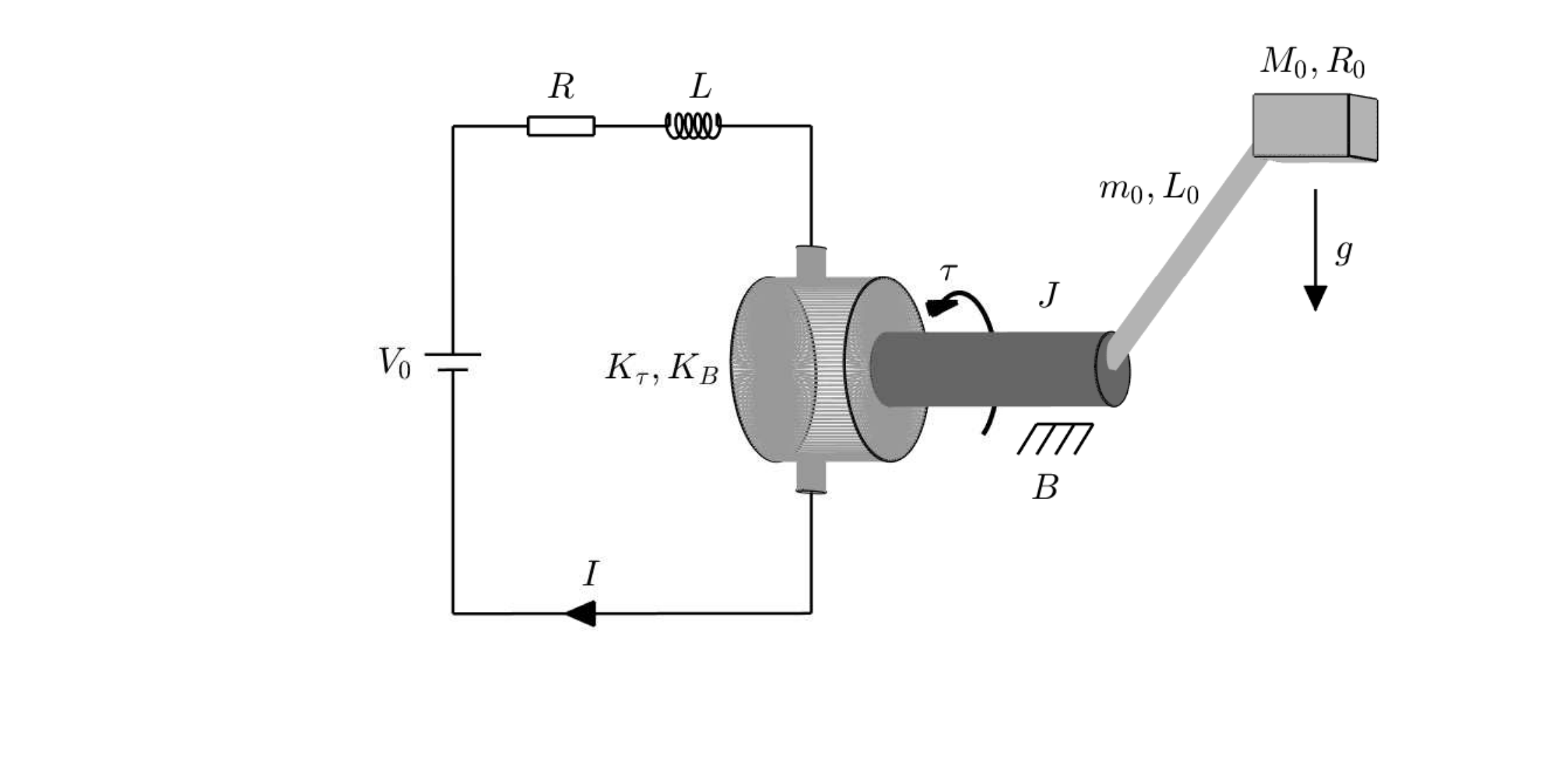}
  \caption{Schematic of the electromechanical system.}
  \label{fig1}
\end{figure}

Consider a electromechanical system \cite{Li2014Tong} shown in Fig. \ref{fig1},
whose dynamics is described as
\begin{align}
\label{equ47}
\begin{cases}
M\ddot{q}+B\dot{q}+N\sin(q)=I,\\
L\dot{I}+K_B\dot{q}+RI=V_\varepsilon,
\end{cases}
\end{align}
where $q(t)$, $I(t)$ and $V_\varepsilon(t)$ are the angular motor position, the motor armature current, and the input control voltage, respectively, $M=\frac{J}{K_\tau}+\frac{m_0L_0^2}{3K_\tau}+\frac{M_0L_0^2}{K_\tau}+\frac{2M_0R_0^2}{5K_\tau}$, $N=\frac{m_0L_0g}{2K_\tau}+\frac{M_0L_0g}{K_\tau}$, $B=\frac{B_0}{K_\tau}$, whose meanings and values of system symbols are shown in Table \ref{tab1}. 
\begin{table*}
\centering
\caption{}
\begin{tabular}{ccc}
\toprule
Symbol&Meaning&Value\\
\midrule
$J$&the rotor inertia&$1.625\times 10^{-3}{\rm kg\cdot m^2}$\\
$m_0$&the link mass&$0.506 {\rm kg}$\\
$M_0$&the load mass&$0.434 {\rm kg}$\\
$L_0$&the link length&$0.305 {\rm m}$\\
$R_0$&the radius of the load&$0.023 {\rm m}$\\
$B_0$&the coefficient of viscous friction at the joint&$16.25\times 10^{-3} {\rm N\cdot m\cdot s/rad}$\\
$L$&the armature inductance&$15{\rm H}$\\
$R$&the armature resistance&$5.0{\rm \Omega}$\\
$K_\tau$&the electromechanical conversion coefficient of armature current to torque&$0.90{\rm N\cdot m/A}$\\
$K_B$&the back EMF coefficient&$0.90{\rm N\cdot m/A}$\\
$g$&the gravity coefficient&$9.81 {\rm N/s^2}$\\
\bottomrule
\end{tabular}
\label{tab1}
\end{table*}

Let $x_1(t)=q(t),x_2(t)=\dot{q}(t),x_3(t)=\frac{I(t)}{M}$ and $u(t)=\frac{V_\varepsilon(t)}{ML}$. Then, system \eqref{equ47} with  mismatched uncertainties can be written as
\begin{align}
\label{equ48}
\begin{cases}
\dot{x}_1(t)=x_2(t)+\omega_1(t),\\
\dot{x}_2(t)=x_3(t)-\frac{N}{M}\sin(x_1(t))-\frac{B}{M}x_2(t)+\omega_2(t),\\
\dot{x}_3(t)=u(t)-\frac{K_b}{ML}x_2(t)-\frac{R}{ML}x_3(t)+\omega_3(t).
\end{cases}
\end{align}
Therefore, it follows from Eq. \eqref{equ48}, Assumptions \ref{assum2} and \ref{assum4} that one can choose $\underline{g_i}=0.1$, $\bar{g}_i=10$ $(i=1,2,3)$ and $L_1=1$, $L_2=\frac{N+B}{M}$, $L_3=\frac{K_B+R}{ML}$. In simulations, the disturbances $\omega_1(t)=2\sin(5t)$, $\omega_2(t)=5\cos(2t)$, $\omega_3(t)=10\sin(t)$ and the reference signal $y_r(t)=\sin(10t)+2$.

\begin{figure}[htbp]
\centering
\begin{minipage}[t]{0.5\textwidth}
\centering
\subfloat[][The tracking error and performance bounds.]{\label{fig2a}\includegraphics[width=8cm,height=5cm]{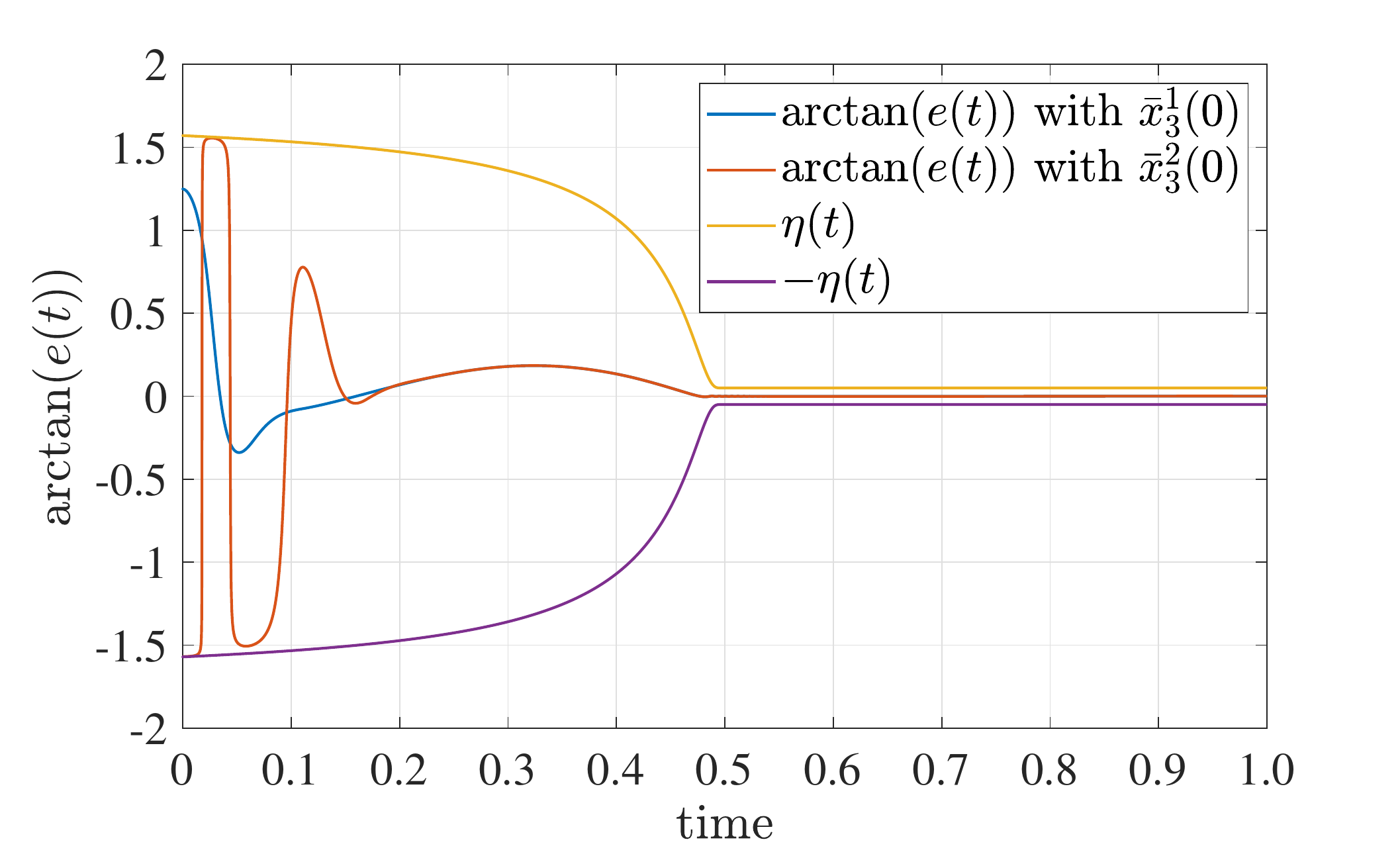}}
\end{minipage}
\begin{minipage}[t]{0.5\textwidth}
\centering
\subfloat[][The reference signal and output trajectories.]{\label{fig2b}\includegraphics[width=8cm,height=5cm]{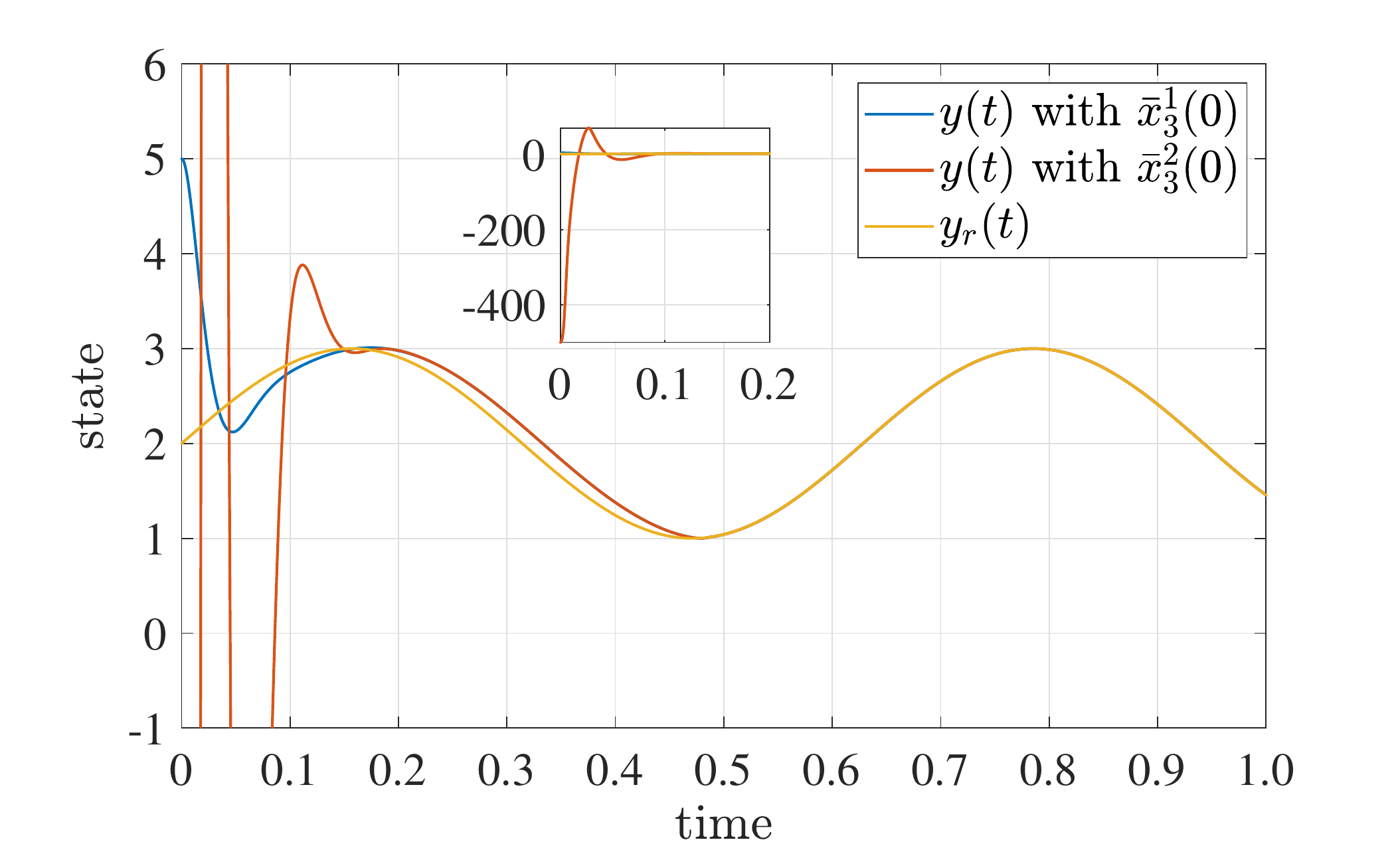}}
\end{minipage}
\caption{State performances under controllers with approximating structures in Theorem \ref{theo1}.}
\label{fig2}
\end{figure}

\begin{figure}[htbp]
\centering
\begin{minipage}[t]{0.5\textwidth}
\centering
\subfloat[][The tracking error and performance bounds.]{\label{fig3a}\includegraphics[width=8cm,height=5cm]{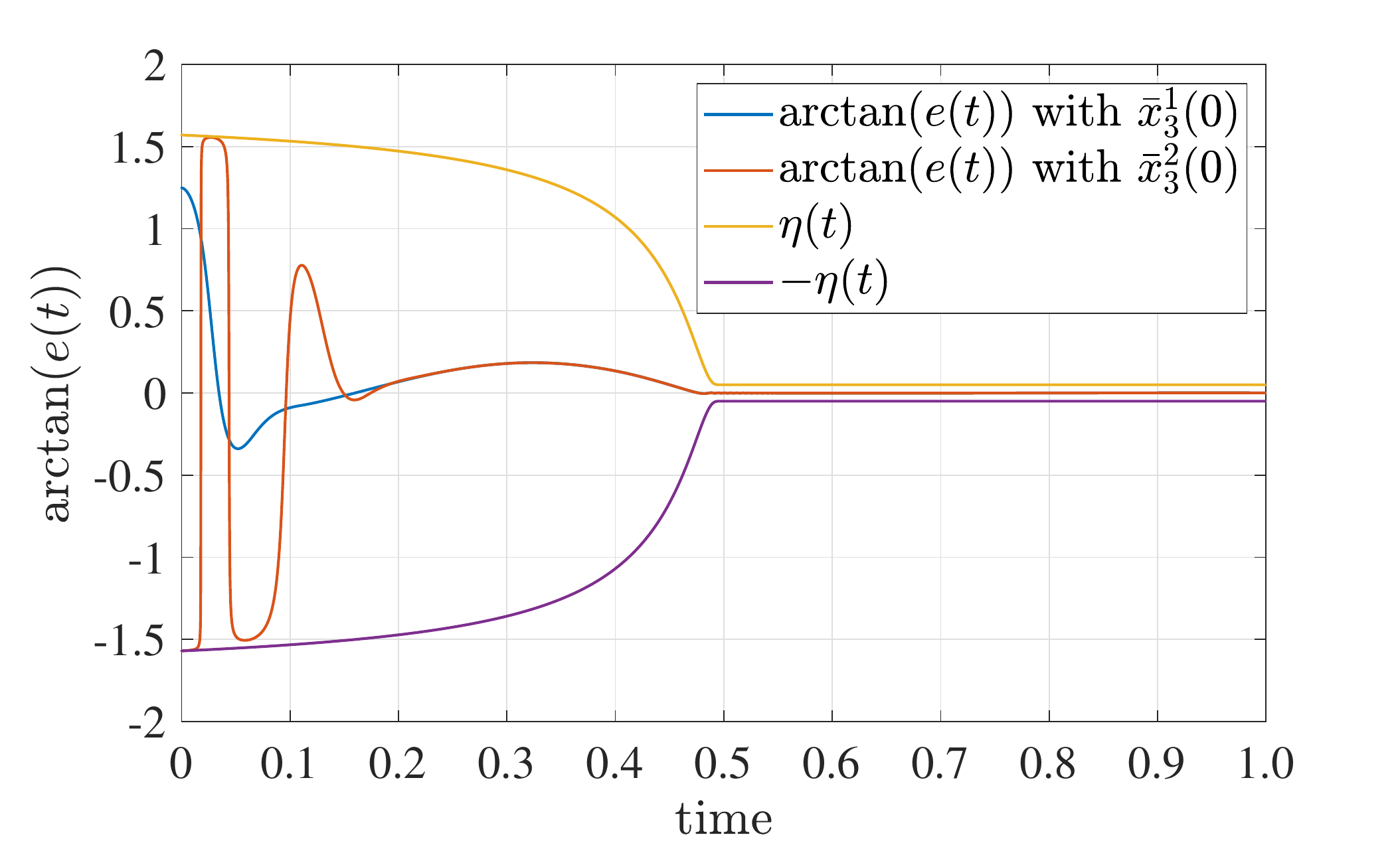}}
\end{minipage}
\begin{minipage}[t]{0.5\textwidth}
\centering
\subfloat[][The reference signal and output trajectories.]{\label{fig3b}\includegraphics[width=8cm,height=5cm]{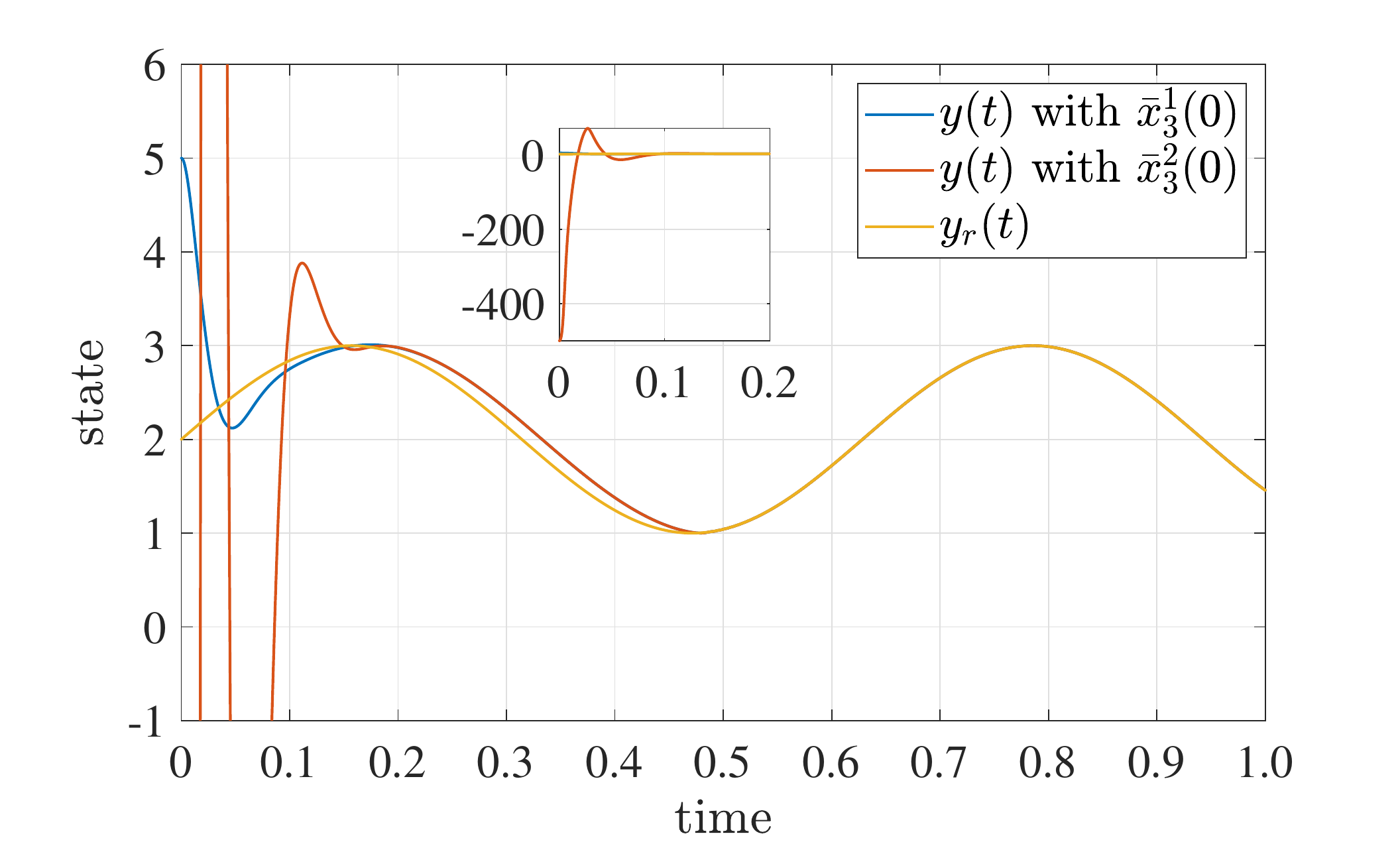}}
\end{minipage}
\caption{State performances under controllers without approximating structures in Theorem \ref{theo2}.}
\label{fig3}
\end{figure}

Take $m=11$ and set the Gaussian membership functions as
\begin{align*}
\mu_{\Omega_i^j}(y_r)=10{\rm e}^{-\frac{(y_r-\upsilon_i^j)^2}{10}},
\end{align*}
where $v_i=(v_i^1,\cdots,v_i^{11})^\top=(-20,\cdots,-4,0,4,\cdots,20)^\top$ for $i=1, 2, 3$. Moreover, control design parameters are selected as
\begin{align*}
b&=0.1, c=0.05, h=1, T=0.5, a=\frac{\frac{\pi}{2}-c}{{\rm e}^{-b}}=1.6807,\\
\delta_1&=\sigma_1=10^{10},\delta_i=\sigma_i=\rho_i=\tau_i=10^{10}~(i=2,3),\\
\varpi_i&=\mu_i=10~(i=1,2),\varrho_i=10,\lambda_i=10^{-5}~(i=2,3),\\
\varpi_3&=5\times 10^3,\mu_3=10,
\end{align*}
and two initial values are set as $\bar{x}_3^1(0)=(5,3,2)^\top$ and $\bar{x}_3^2(0)=-100\bar{x}_3^1(0)$ with $\hat{\theta}_i(0)=\mathbf{0}_{11}$ $(i=1,2,3)$. Figs. \ref{fig2} and \ref{fig3} present the simulation results.

It can be observed from Figs. \ref{fig2a} and \ref{fig3a} that tracking error $e(t)$ always satisfies that $-\eta(t)<{\rm arctan}(e(t))<\eta(t)$, equivalently, $\lvert e(t)\rvert< \tan(\eta(t))$ for any $t\geq0$, and $\lvert e(t)\rvert< \tan(0.05)$ for $t\geq 0.5$. Therefore, the prescribed transient and steady state performances with prescribed time are achieved via proposed controllers. It should be emphasized that the state performances without approximating structures in Fig. \ref{fig3} are almost identical to those with approximating structures in Fig. \ref{fig2}, but the computational complexity without approximating structures is far below the latter. The control mechanism without approximating structures can thus save more computation resources and have more potential to practical application.

\begin{figure}
  \centering
  \includegraphics[width=8cm,height=4cm,trim=30 130 30 20,clip]{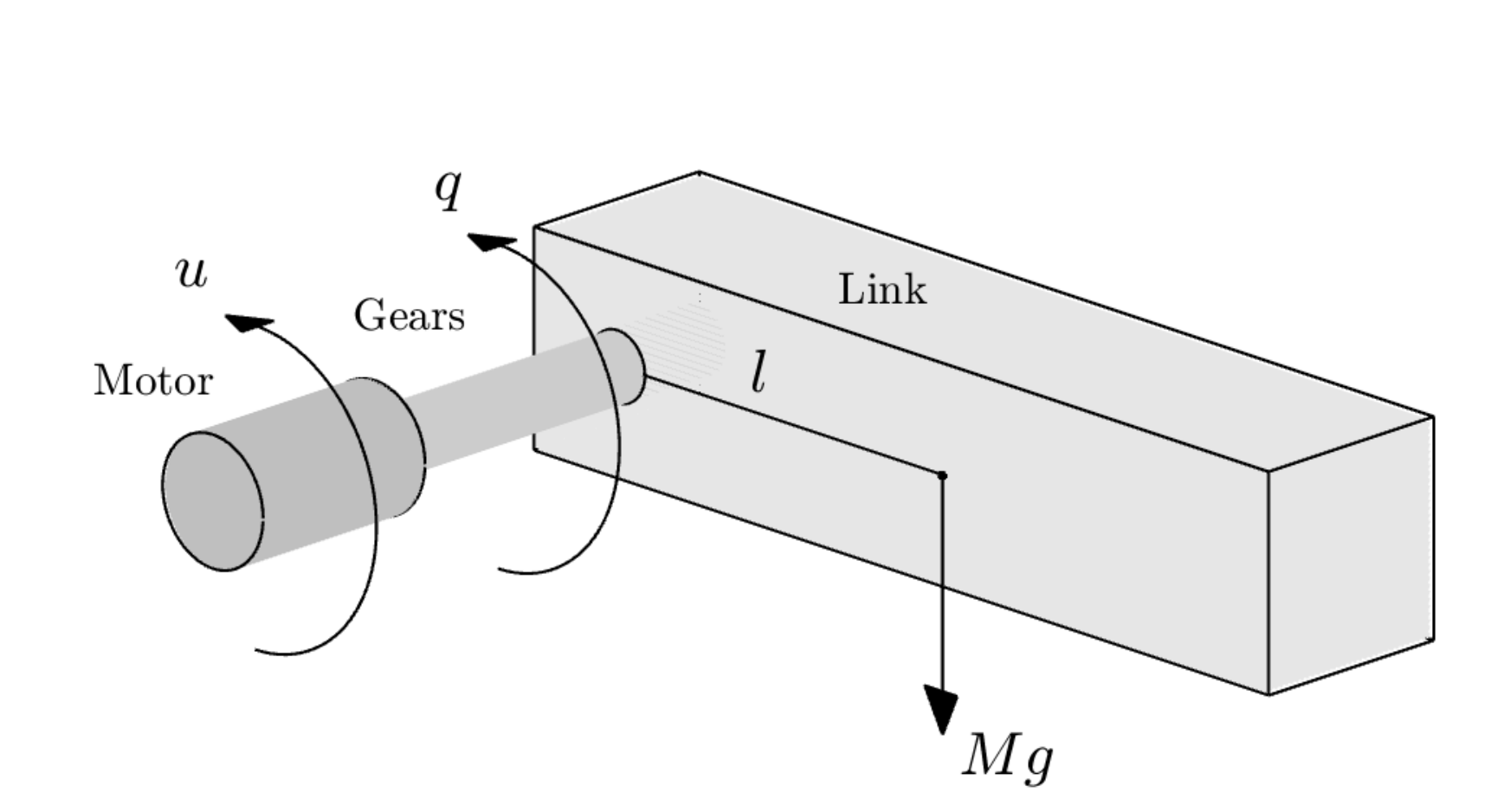}
  \caption{A single-link manipulator.}
  \label{fig4}
\end{figure}

In what follows, a single-link manipulator \cite{Mao2022Wu} shown in Fig. \ref{fig4} is employed to compare the proposed controller without approximating structures and the one designed in \cite{Qiu2022Wang}.
The single-link manipulator's dynamics is 
\begin{align}
\label{equ50}
I\ddot{q}(t)+B\dot{q}(t)+Mgl\sin(q(t))=u(t),
\end{align}
where $g=9.81{\rm N/s^2}$ is the gravity coefficient, $q(t)$ is the angles of the link, $u(t)$ is the control input, $I$ is the total rotational inertias of the link and the motor, $l$ is  the distance between the joint axis and the link center of mass, $M$ is the total mass of the single link, and $B$ is the damping coefficients.  For simplicity, take $I=1{\rm kg\cdot m^2}$, $B=2{\rm kg\cdot m/s}$, $M=1{\rm kg}$ and $l=1{\rm m}$.

\begin{figure}
  \centering
  \includegraphics[width=8cm,height=5cm,trim=0 0 10 10,clip]{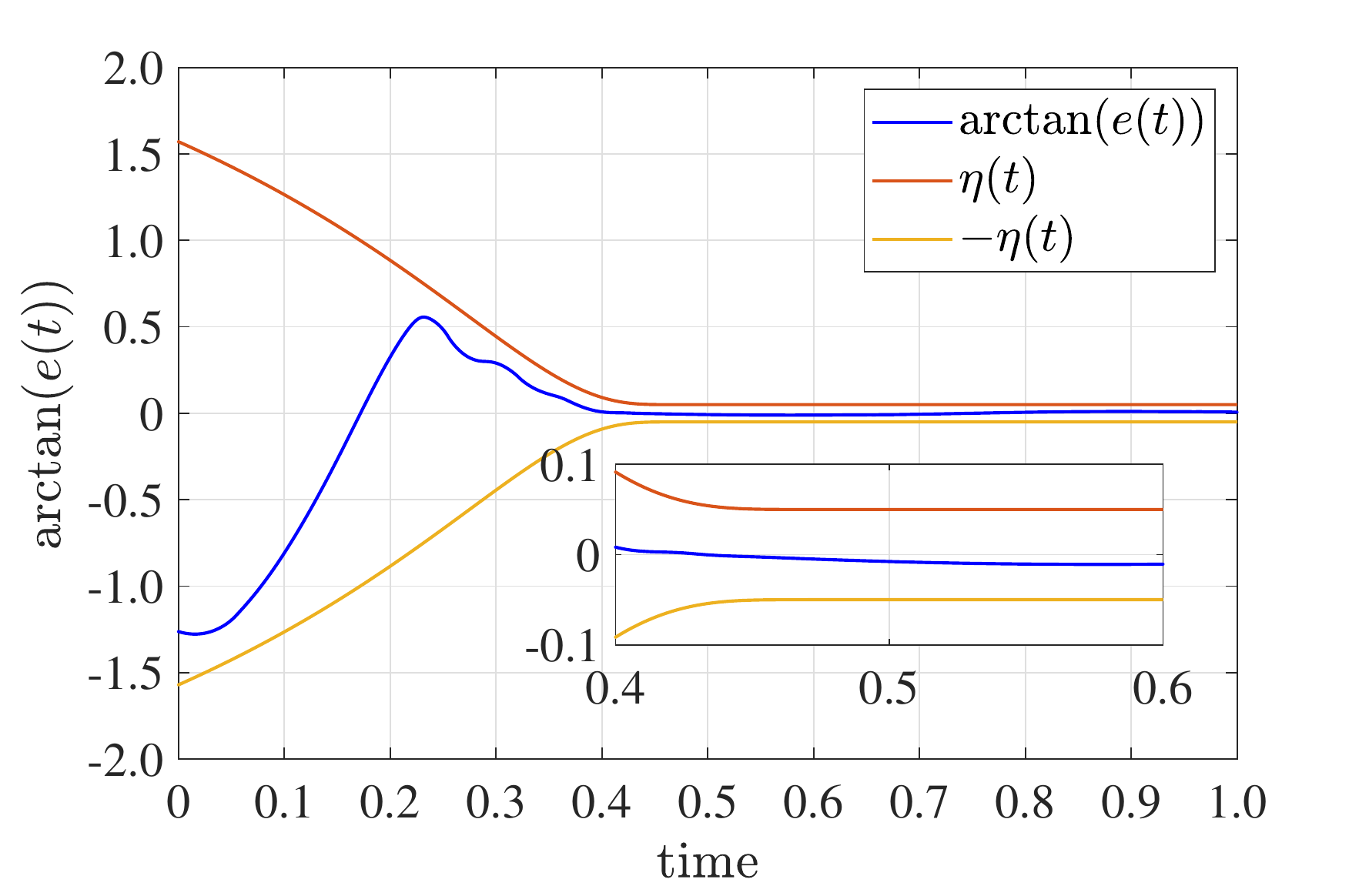}
  \caption{The performance bounds and the tracking error with the proposed controller $u_1(t)$.}
  \label{fig5}
\end{figure}

\begin{figure}
  \centering
  \includegraphics[width=8cm,height=5cm,trim=0 0 10 10,clip]{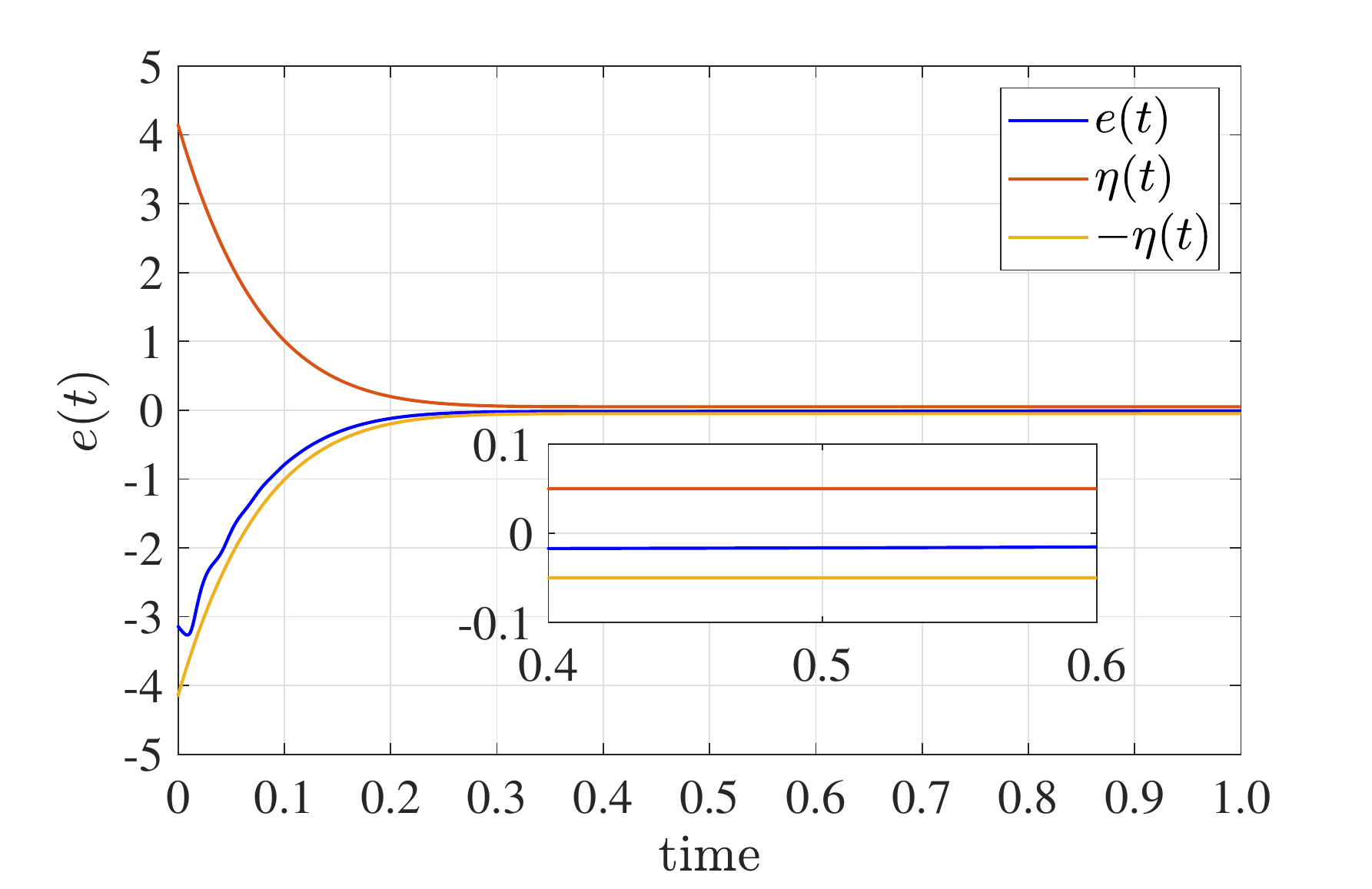}
  \caption{The performance bounds and the tracking error with controller $u_2(t)$ proposed by \cite{Qiu2022Wang}.}
  \label{fig6}
\end{figure}

\begin{figure}
  \centering
  \includegraphics[width=8cm,height=5cm,trim=0 0 10 10,clip]{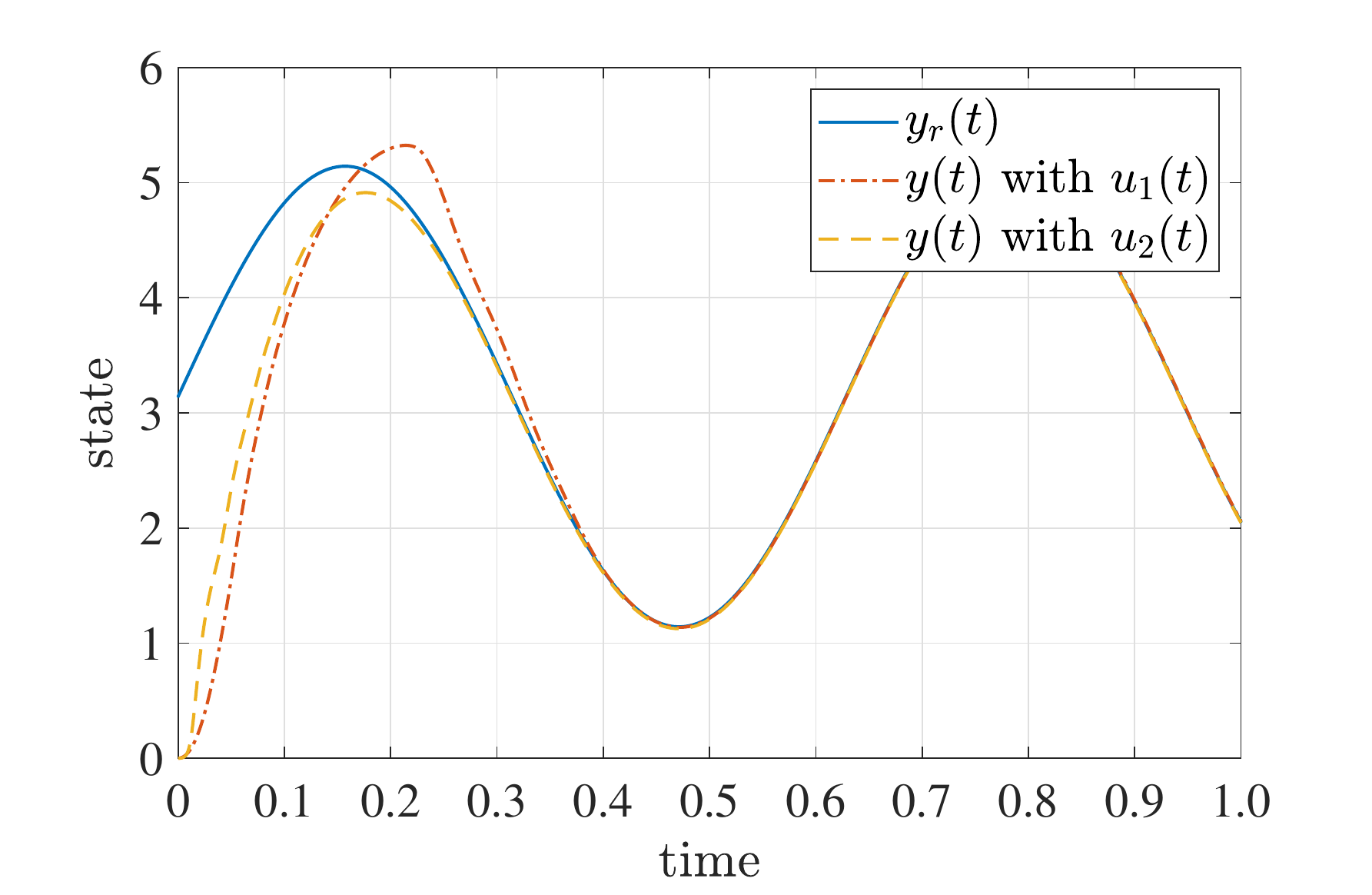}
  \caption{The reference signal and output trajectories.}
  \label{fig7}
\end{figure}

\begin{figure}
  \centering
  \includegraphics[width=8cm,height=5cm,trim=0 0 10 10,clip]{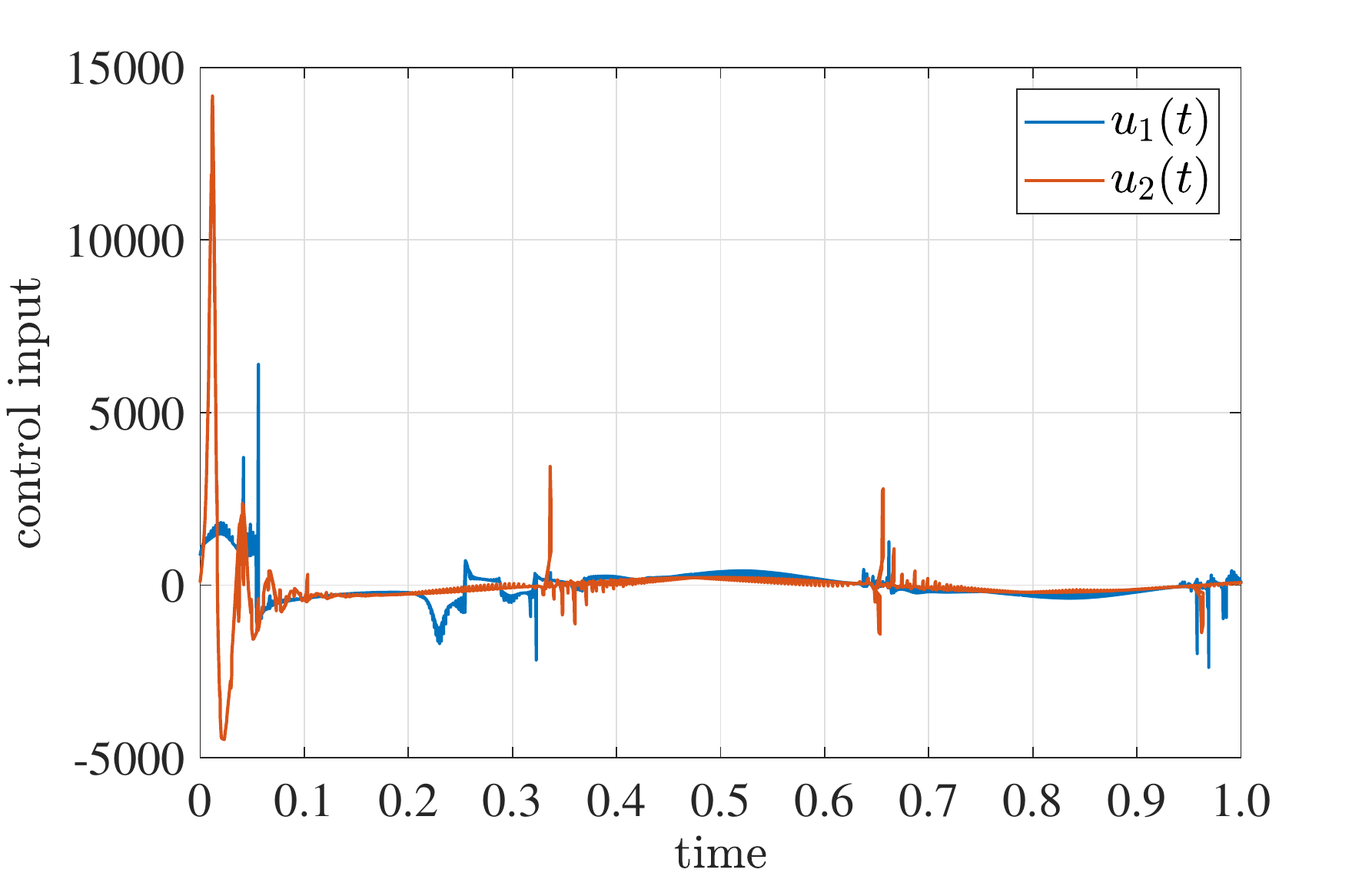}
  \caption{Control inputs.}
  \label{fig8}
\end{figure}

Denote $x_1(t)=q(t)$ and $x_2(t)=\dot{q}(t)$. Then, according to Eq. $(\ref{equ50})$, the dynamics with external disturbance is
\begin{align*}
\begin{cases}
\dot{x}_1(t)=x_2(t)+\omega_1(t),\\
\dot{x}_2(t)=-\frac{1}{I}(Bx_2(t)+Mgl\sin(x_1(t)))+\frac{1}{I}u(t)+\omega_2(t).
\end{cases}
\end{align*}
Let the reference signal $y_r(t)=\pi+2\sin(10t)$, and disturbances $\omega_1(t)=0,\omega_2(t)=10\cos(5t)$. In this example, take $\underline{g}_1=\underline{g}_2=0.5,$ $\bar{g}_1=\bar{g}_2=10$, $L_1=1,L_2=\frac{B+Mgl}{I}$, $\delta_1=\delta_2=\sigma_1=\sigma_2=\rho_2=\tau_2=10^6$, $\varpi_2=\varrho_2=10$ and $\lambda_2=0.001$. The parameters in prescribed performance functions are chosen as $b=0.9,c=0.05,T=0.5,h=1$ and $a=4.1340$. In addition, Gaussian membership functions are the same as above. All initial values are chosen as zero. For the purpose of rigour in simulation, the settling time $t_r$ and initial value of prescribed performance function $\eta(0)$ in \cite{Qiu2022Wang} are set as $t_r=T$ and $\eta(0)=\lvert e(0)\rvert+1$, and the other parameters are the same as those in \cite{Qiu2022Wang}. Therefore, $\psi=\frac{\eta_0^\ell}{\ell t_r}=16.1466$, where $\ell =\frac{2}{13},\eta_0=\eta(0)-\eta_{t_r}$ and $\eta_{t_r}=\tan(c)$. For convergence, denote the proposed controller and the one in \cite{Qiu2022Wang} by $u_1(t)$ and $u_2(t)$, respectively. The simulation results are displayed in Figs. \ref{fig5}, \ref{fig6}, \ref{fig7} and \ref{fig8}.

As shown in Figs. \ref{fig5}, \ref{fig6} and \ref{fig7}, both controllers $u_1(t)$ and $u_2(t)$ can guarantee the prescribed transient and steady state performance of the tracking error within prescribed time, which again verifies the validity of Theorem \ref{theo2}. In addition, it can be seen from Fig. \ref{fig8} that the absolute value of the proposed controller $u_1(t)$ with prescribed-time prescribed performance is generally less than controller $u_2(t)$ with finite-time prescribed performance, which demonstrates the effectiveness and practicality of the proposed methods.

\section{Conclusion}
\label{sec5}
In this paper, the adaptive fuzzy tracking control with global prescribed-time prescribed performance for strict-feedback nonlinear systems with mismatched uncertainties has been investigated. Firstly, a class of prescribed-time prescribed performance functions independent of initial values and an error transformation function are designed. Secondly, two adaptive fuzzy controllers with and without approximating structures are designed to guarantee prescribed-time prescribed performance of the tracking error and the global uniform boundedness of all closed-loop signals. With a novel Lyapunov-like energy function, the differential explosion problem frequently occurring in backstepping techniques is solved.  It is worth noting that no Nussbaum-type functions are used and no singular phenomenon occurs in the control design, and thus complex calculations can be avoided. Finally, some practical examples are employed to demonstrate the validity and effectiveness of the proposed methods. In future studies, the focus will be on the leader-following consensus with prescribed-time prescribed performance for strict-feedback multi-agent systems owing to their broad applications in various fields.
\section*{Acknowledgments}
This work is supported by the National Natural Science Foundation of China under Grants 61973241 and 62176127.

\vspace{-10 mm}
\begin{IEEEbiography}[{\includegraphics[width=1in,height=1.25in,clip,keepaspectratio]{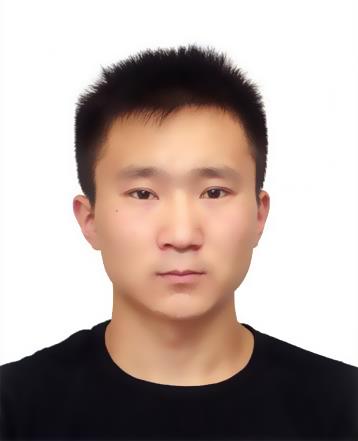}}]{Bing Mao}
received the B.Sc. degree in mathematics from Wuhan University, Wuhan, China, in 2019.
He is currently pursuing the Ph.D. degree with the School of Mathematics and Statistics, Wuhan University, Wuhan, China.
His current research interests include topology identification of complex networks, and control of multi-agent systems.
\end{IEEEbiography}

\vspace{-10 mm}
\begin{IEEEbiography}[{\includegraphics[width=1in,height=1.25in,clip,keepaspectratio]{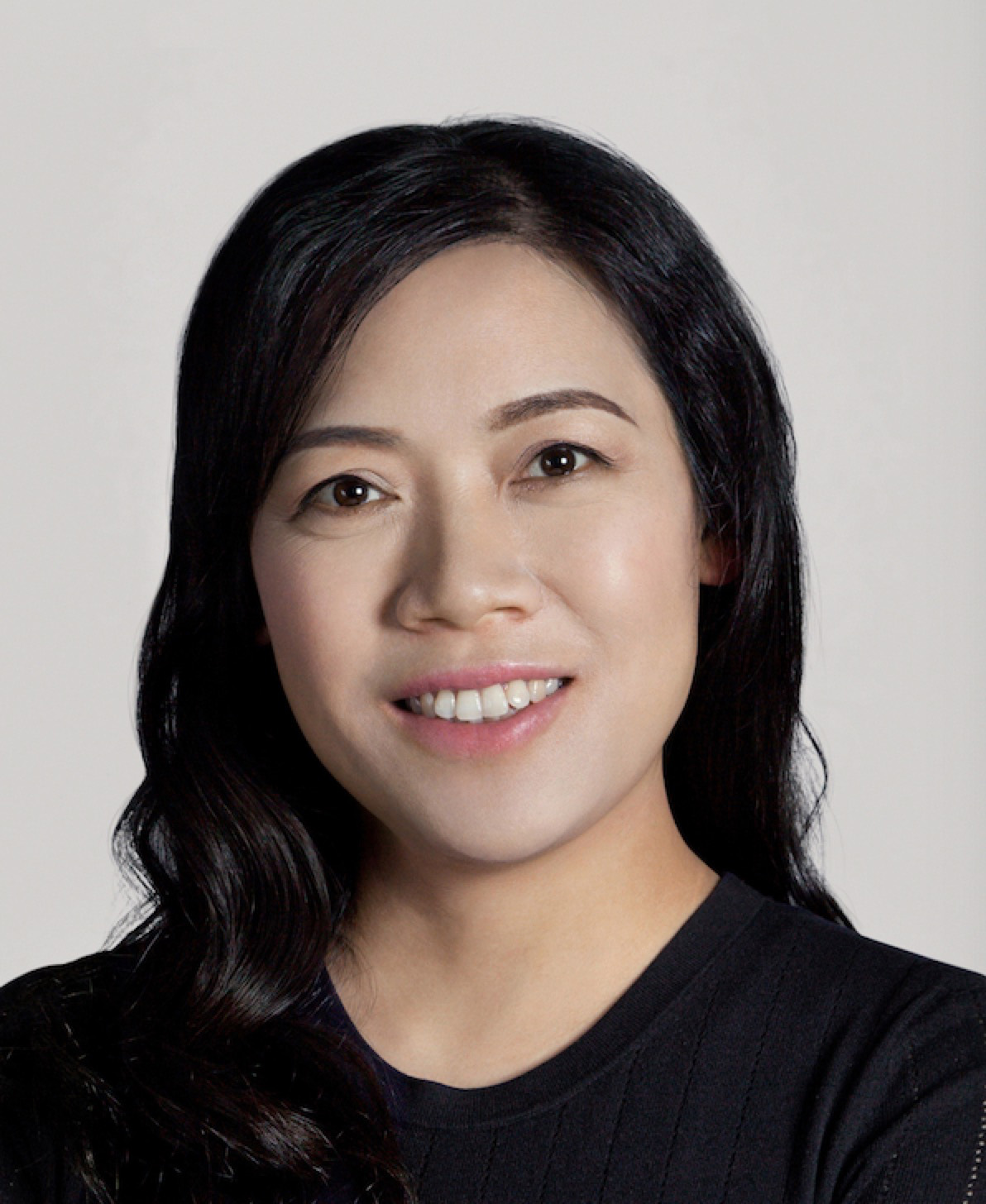}}]{Xiaoqun Wu}
received the B.Sc. degree in applied mathematics and the Ph.D. degree in computational mathematics from Wuhan University, Wuhan, China, in 2000 and 2005, respectively.
\par
She is currently a Professor with the School of Mathematics and Statistics, Wuhan University. She held several visiting positions in Hong Kong, Australia and America over the last few years. Her current research interests include complex networks, nonlinear dynamics, and chaos control. She has published over 70 SCI journal papers in these areas.
\par
Prof. Wu was a recipient of the Second Prize of the Natural Science Award from the Hubei Province, China in 2006, the First Prize of the Natural Science Award from the Ministry of Education of China in 2007, and the First Prize of the Natural Science Award from the Hubei Province, China, in 2013.  In 2017, she was awarded  the 14th Chinese Young Women Scientists Fellowship and the Natural Science Fund for Distinguished Young Scholars of Hubei Province.   She is serving as an Associate Editor for IEEE Transactions on Circuits and Systems II.
\end{IEEEbiography}

\vspace{-10 mm}
\begin{IEEEbiography}[{\includegraphics[width=1in,height=1.25in,clip,keepaspectratio]{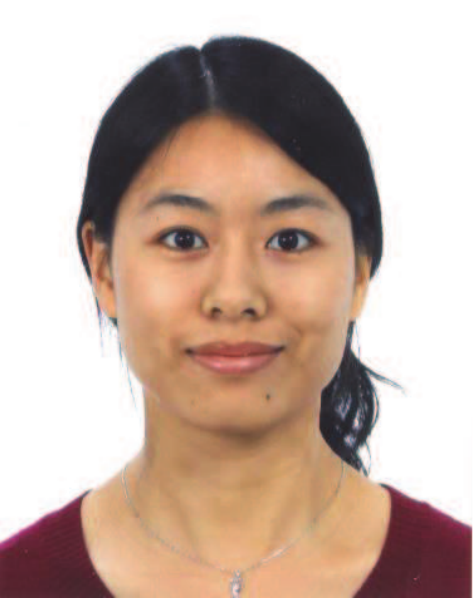}}]{Hui Liu}
 (S'11-M'14) received the B.S. and Ph.D. degrees in computational mathematics from Wuhan University, China, in 2005 and 2010 respectively. And she also received the Ph.D. degree in the field of systems and control from the University of Groningen, the Netherlands, in 2013.
She is currently an associate professor with the School of Artificial Intelligence and  Automation, Huazhong University of Science and Technology, Wuhan, China. She held visiting positions with the Academy of Mathematics and Systems Science, Chinese Academy of Sciences, Beijing, China, and with the Department of Electronic and Information Engineering, the Hong Kong Polytechnic University, Hong Kong. Her main research interests are in complex networks, cooperative control of networked systems, and intelligent control systems.
\end{IEEEbiography}

\vspace{-10 mm}
\begin{IEEEbiography}[{\includegraphics[width=1in,height=1.25in,clip,keepaspectratio]{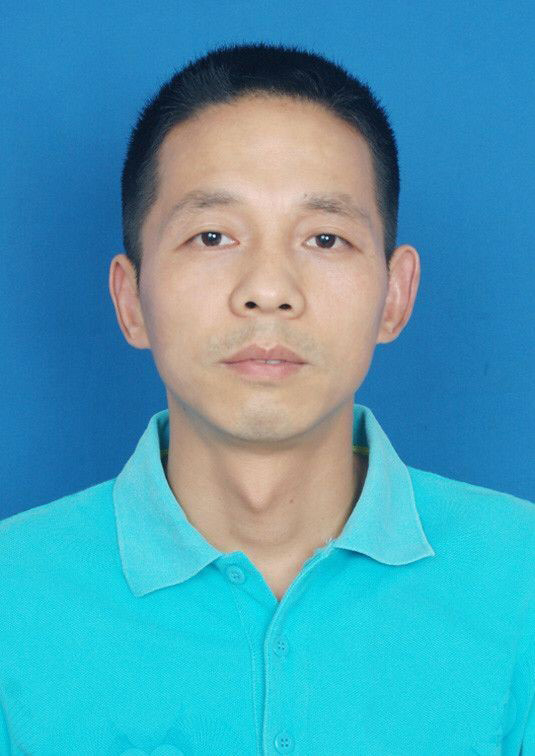}}]{Yuhua Xu}
received his Ph.D. degree in control theory and control engineering from Donghua University, China, in 2011. From 2012 to 2014 he was a postdoctoral fellow in School of Computing at Wuhan University, Wuhan, China.
\par
Currently he is a Professor at School of Finance, Nanjing Audit University, Jiangsu 211815, China. His current research interests include complex networks, nonlinear dynamics, nonlinear finance systems and chaos control. He has published over 50 SCI journal papers in these areas. Prof. Xu was a recipient of the Excellent Young Key Talent Plan in Hubei Province, China, in 2011. The Young and middle-aged academic leaders of “Qing-Lan Engineering” in Jiangsu Province, China, in 2017. The Talent Plan of “Six Talents Peaks Project” in Jiangsu Province, China, in 2018.
\end{IEEEbiography}

\vspace{-10 mm}
\begin{IEEEbiography}[{\includegraphics[width=1in,height=1.25in,clip,keepaspectratio]{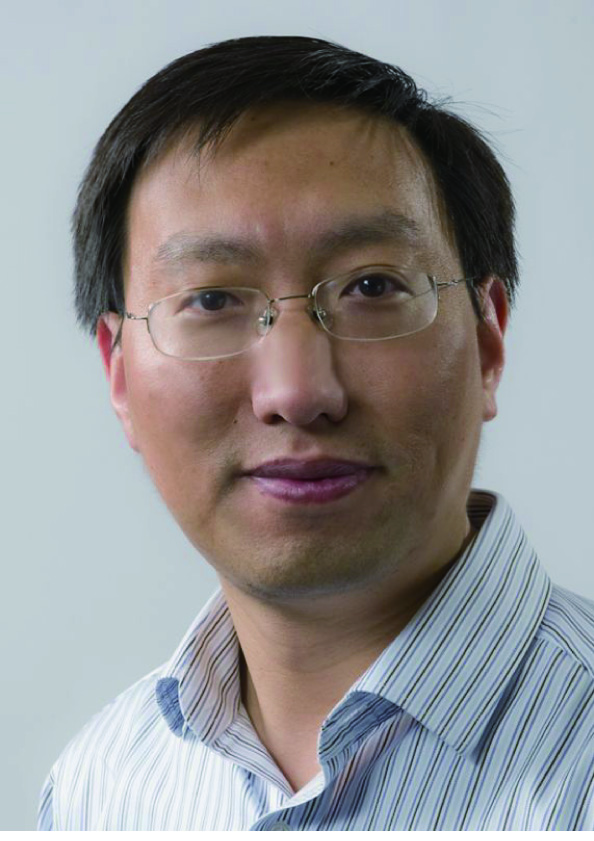}}]{Jinhu L\"{u}}
(M'03-SM'06-F'13) received the Ph.D. degree in applied mathematics from the Academy of Mathematics and Systems Science, Chinese Academy of Sciences, Beijing, China, in 2002.
\par
He was a Professor with RMIT University, Melbourne, VIC, Australia, and a Visiting Fellow with Princeton University, Princeton, NJ, USA. Currently, he is the Dean with the School of Automation Science and Electrical Engineering, Beihang University, Beijing, China. He is also a Professor with the AMSS, Chinese Academy of Sciences. He is a Chief Scientist of the National Key Research and Development Program of China and a Leading Scientist of Innovative Research Groups of the National Natural Science Foundation of China. His current research interests include complex networks, industrial Internet, network dynamics and cooperation control.
\par
Dr. L\"{u} was a recipient of the prestigious Ho Leung Ho Lee Foundation Award in 2015, the National Innovation Competition Award in 2020, the State Natural Science Award three times from the Chinese Government in 2008, 2012, and 2016, respectively, the Australian Research Council Future Fellowships Award in 2009, the National Natural Science Fund for Distinguished Young Scholars Award, and the Highly Cited Researcher Award in engineering from 2014 to 2019. He is/was an Editor in various ranks for 15 SCI journals, including the Co-Editor-in-Chief of IEEE TII. He served as a member in the Fellows Evaluating Committee of the IEEE CASS, the IEEE CIS, and the IEEE IES. He was the General Co-Chair of IECON 2017. He is the Fellow of IEEE and CAA.
\end{IEEEbiography}


\end{document}